\documentclass[11pt]{article}       
\usepackage{graphicx}

\usepackage{tkz-euclide}

\usepackage[hyphens]{url}
\RequirePackage[OT1]{fontenc}
\RequirePackage{amsthm,amsmath, bbm, amsfonts, amssymb, graphics, graphicx, subfigure, color, multirow}

\RequirePackage{natbib}
\RequirePackage[colorlinks,citecolor=blue,urlcolor=blue,breaklinks]{hyperref}

\usepackage{algpseudocode,subfigure}
\usepackage{algorithm}
\usepackage{comment}

\usepackage{pstricks-add}
\usepackage[draft]{todonotes}

\usepackage{graphicx}
\usepackage{epstopdf}
\usepackage[singlelinecheck=false]{caption}

\usepackage{algpseudocode}
\usepackage{comment}

\newtheorem{exam}{Example}[]

\newtheorem{remark}{Remark}
\newtheorem{theorem}{Theorem}
\newtheorem{lemma}{Lemma}

\newcommand{\iid}{\stackrel{\mathrm{iid}}{\sim}}
\newcommand{\ind}{\stackrel{\mathrm{ind}}{\sim}}

\newcommand{\mfdr}{m\textsc{fdr}}
\newcommand{\mfnr}{m\textsc{fnr}}
\newcommand{\fdr}{\textsc{fdr}}
\newcommand{\fnr}{\textsc{fnr}}
\newcommand{\mfdrspace}{m\textsc{fdr }}
\newcommand{\mfnrspace}{m\textsc{fnr }}
\newcommand{\fdrspace}{\textsc{fdr }}
\newcommand{\fnrspace}{\textsc{fnr }}
\newcommand{\vx}{\boldsymbol{x}}
\newcommand{\bdelta}{\boldsymbol{\delta}}

\usepackage{multirow}
\usepackage{rotating}
\usepackage{epstopdf}
\usepackage{color}

\title{Where to find needles in a haystack?
}

\author{Zhigen Zhao 
\\
Department of Statistical Science\\Temple University\\ Philadelphia, PA 19122\\ USA
}
\date{}

\begin{document}
\maketitle

  In many existing methods of multiple comparison, one starts with either Fisher's p-value or the local fdr. One commonly used p-value, defined as the tail probability exceeding the observed test statistic under the null distribution, fails to use information from the distribution under the alternative hypothesis. The targeted region of signals could be wrong when the likelihood ratio is not monotone. The oracle local fdr based approaches could be optimal because they use the probability density functions of the test statistic under both the null and alternative hypotheses. However, the data-driven version could be problematic because of the difficulty and challenge of probability density function estimation. In this paper, we propose a new method, {\bf C}df and {\bf L}ocal fdr {\bf A}ssisted multiple {\bf T}esting method (CLAT), which is optimal for cases when the p-value based methods are optimal and for some other cases when p-value based methods are not. Additionally, CLAT only relies on the empirical distribution function which quickly converges to the oracle one. Both the simulations and real data analysis demonstrate the superior performance of the CLAT method. Furthermore, the computation is instantaneous based on a novel algorithm and is scalable to large data sets.


\section{Introduction}\label{sec:intro}

In modern scientific investigations, scientists often need to make statistical inferences for thousands or even millions of parameters simultaneously when conducting their research.
A tremendous increase in statistical methodologies, some of which are impressively creative, have been proposed to deal with various related issues.
In this paper, we focus on large-scale simultaneous hypothesis testing, or large scale multiple comparison procedures(MCP). Namely, we test a collection of $n$ hypotheses: 
\begin{equation}\label{eqn:hypothesis}
H_{0,i}  \qquad  \mbox{vs.}  \qquad H_{1,i}, \qquad \qquad i = 1,2, \ldots, n.
\end{equation}
Associated with these hypotheses is a collection of test  statistics $X_1,X_2,\cdots,X_n$.

\subsection{Model and Error Rates}\label{sec:model}

For $i=1,2,\cdots,n$, assume that the test statistic $X_i\sim f_0(x)$ under the null hypothesis $H_{0i}$ and $X_i\sim f_1(x)$ under the alternative hypothesis $H_{1i}$ where $f_0(x)$ and $f_1(x)$ are two probability density functions. Let $\pi_1$ be the proportion of non-true nulls. We consider the following two-group model (\cite{Efron:2008, Efron:2010b})
\begin{equation}\label{model}
X_i\iid (1-\pi_1)f_0(x)+\pi_1 f_1(x).
\end{equation}
Similarly, let $F_0(x)$ and $F_1(x)$ be the cumulative distribution functions of $X_i$'s under the null and alternative hypotheses respectively. Then the cumulative distribution function of the $X_i$'s is $F(x)=(1-\pi_1)F_0(x)+\pi_1 F_1(x)$. 

Model (\ref{model}) has a natural connection to the following hierarchical model. Let $\theta_i$ be the indicator that the $i$-th hypothesis $H_{0,i}$ is false. Assume that
\begin{eqnarray*}
	\left\{ \begin{array}{c}
		X_i|\theta_i \ind (1-\theta_i)f_0(x_i) + \theta_i f_1(x_i),\\
		\theta_i \iid Bernoulli(\pi_1).
		\end{array}
		\right.
	\end{eqnarray*}

For any given test statistics $X_i$'s, let $\bdelta\in \{0,1\}^n$ be the decision based on a certain procedure. Here $\delta_i=1$ means that the $i$-th hypothesis is rejected. Define
\[
\fdr=E\frac{\sum_i(1-\theta_i)\delta_i}{\sum_i\delta_i\vee 1}, \fnr = E\frac{\sum_i\theta_i(1-\delta_i)}{\sum_i(1-\delta_i)\vee 1}.
\]
The marginal \fdrspace (\mfdr) and marginal \fnrspace (\mfnr) are defined as
\[
\mfdr=\frac{E\sum_i(1-\theta_i)\delta_i}{E \sum_i\delta_i}, \mfnr = \frac{E\sum_i\theta_i(1-\delta_i)}{E\sum_i(1-\delta_i)}.
\]
It is shown in \cite{Genovese:Wasserman:2002}, that $\mfdr=\fdr + O(n^{-1/2})$. In the testing framework, we are looking for an "optimal" method that minimizes the \mfnrspace subject to a control of \mfdrspace at a designated level, say $q$.



\subsection{Revisit the p-value}\label{sec:pvalue}
The $P$-value, the probability of obtaining a test statistics at least as extreme as the one that was actually observed given that the null hypothesis is true, is defined by Ronald A. Fisher in his research papers and various editions of his influential texts, such as \cite{Fisher:1925} and \cite{Fisher:1935}. 
A small p-value indicates that ``Either an exceptionally rare chance has occurred or the theory is not true'' (\cite{Fisher:1959}, p.39).


In practice, the evidence against the null distribution usually appears on the tail. Thus, a widely used p-value is
\begin{equation}\label{EqDefinep}
p_i=P(X>x_i|H_0)= 1- F_0(x_i),
\end{equation}
where here we assume that large $X$-values support the alternative hypothesis against the null hypothesis. This is the starting point for many testing methods, including the famous BH method (\cite{Benjamini:Hochberg:1995}).

\begin{algorithm}[H]
  \caption{BH Method.  \label{alg:bh}}
  \begin{algorithmic}[1]
    \vspace*{1mm}
    \item Order the p-values as $p_{(1)}\le p_{(2)}\le \cdots\le p_{(n)}$;
    \item Let $R=max_{1\le i\le n}\{i: p_{(i)}\le \frac{i\alpha}{n}\}$;
    \item Reject $H_{i0}$ if and only if $p_i \le p_{(R)}$.
  \end{algorithmic}
\end{algorithm}
Equivalently, a threshold $T=F_0^{-1}(1-p_{(R)})$ is chosen and the $i$-th hypothesis is rejected when the test statistic $X_i$ is greater than or equal to $T$. In other words, we 

\begin{eqnarray}\label{EqThreshold02}
  \begin{array}{rl}
    \textrm{  Reject $H_{0i}$},  & \mbox{if $X_i \geq T$},     \\
    \textrm{ Fail-to-reject $H_{0i}$},     &  \mbox{if   $X_i < T$}.
  \end{array}
\end{eqnarray}

The commonly used p-value given in Equation (\ref{EqDefinep}) does not depend on $f_1(x)$ ($F_1(x)$).
For cases when the likelihood ratio $\Lambda(x)(=\frac{f_1(x)}{f_0(x)})$ is monotone increasing with respect to $x$, known as the monotone likelihood ratio property (MLR) (\cite{Karlin:Rubin:1956a, Karlin:Rubin:1956b}), a small p-value, or equivalently a large test statistic implies stronger evidence against the null.
However, for cases when the MLR does not hold, the belief that ``the larger the $X_i$, the stronger evidence against the null'' is shattered into pieces. In this scenario, a small p-value would favor the null hypothesis  rather than the alternative. 
For example, let $f_0(x_i)=\phi(x_i)$ and $f_1(x_i)=\frac{1}{\sigma}\phi(\frac{x_i-\mu}{\sigma})$ where $\phi(x_i)$ is the probability
density function of the standard normal distribution. If $\sigma<1$, an extremely large observation is more likely to have been generated from the null distribution.
The decision defined  in  (\ref{EqThreshold02}) is no longer appropriate no matter what threshold is picked. Instead, it is more appropriate to consider the following decision:
\begin{eqnarray}\label{EqThreshold04}
  \begin{array}{rl}
    \textrm{ Reject $H_{0i}$},  & \mbox{if  $T \leq  X_i  \leq S $},     \\
    \textrm{ Fail-to-reject $H_{0i}$ },     &  \mbox{otherwise}.
  \end{array}
\end{eqnarray}

\begin{figure}[!htbp]
\centering
\subfigure[]{\label{fig:normal}\includegraphics[width=50mm, height=50mm]{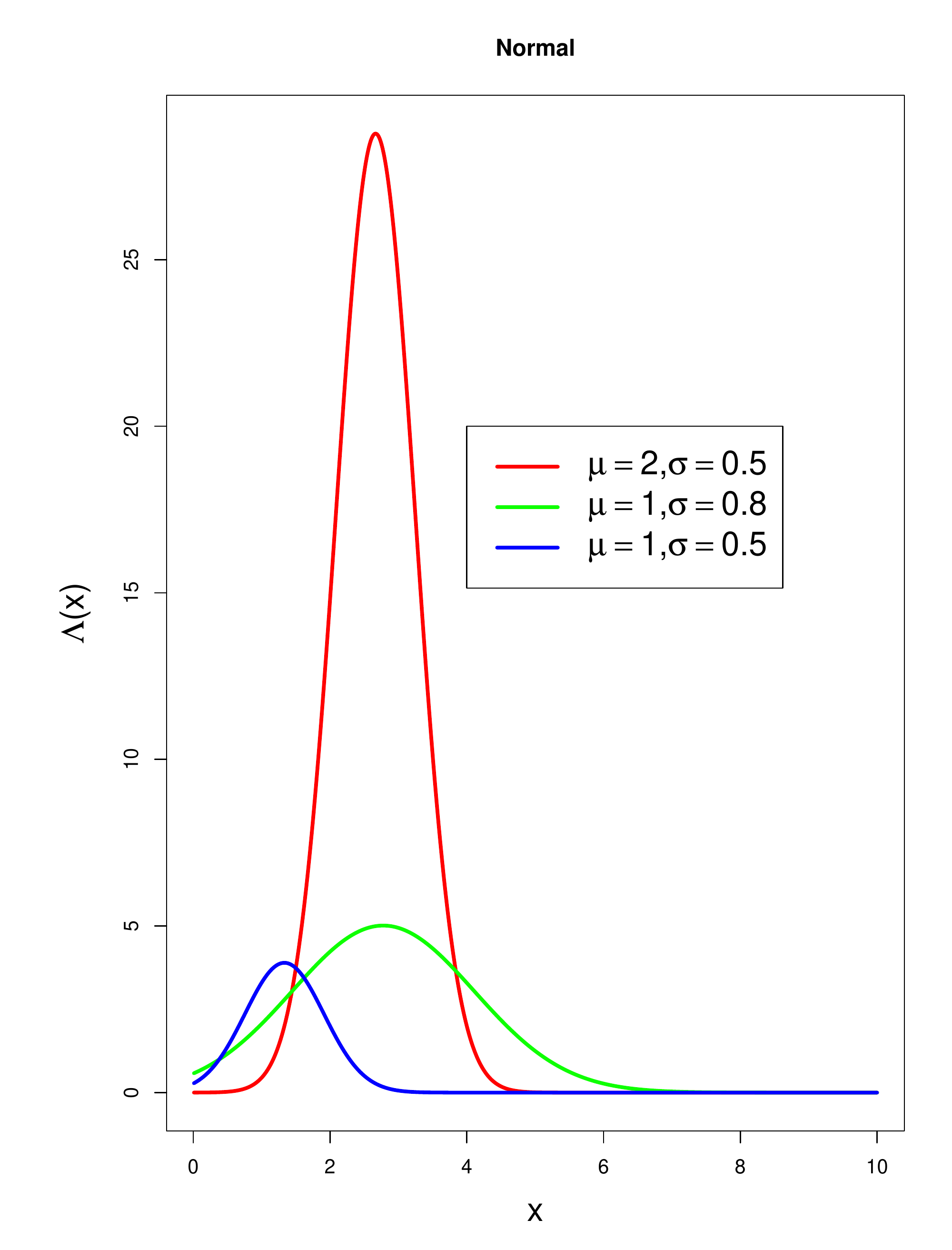}}    
\subfigure[]{\label{fig:gennormal}\includegraphics[width=50mm,height=50mm]{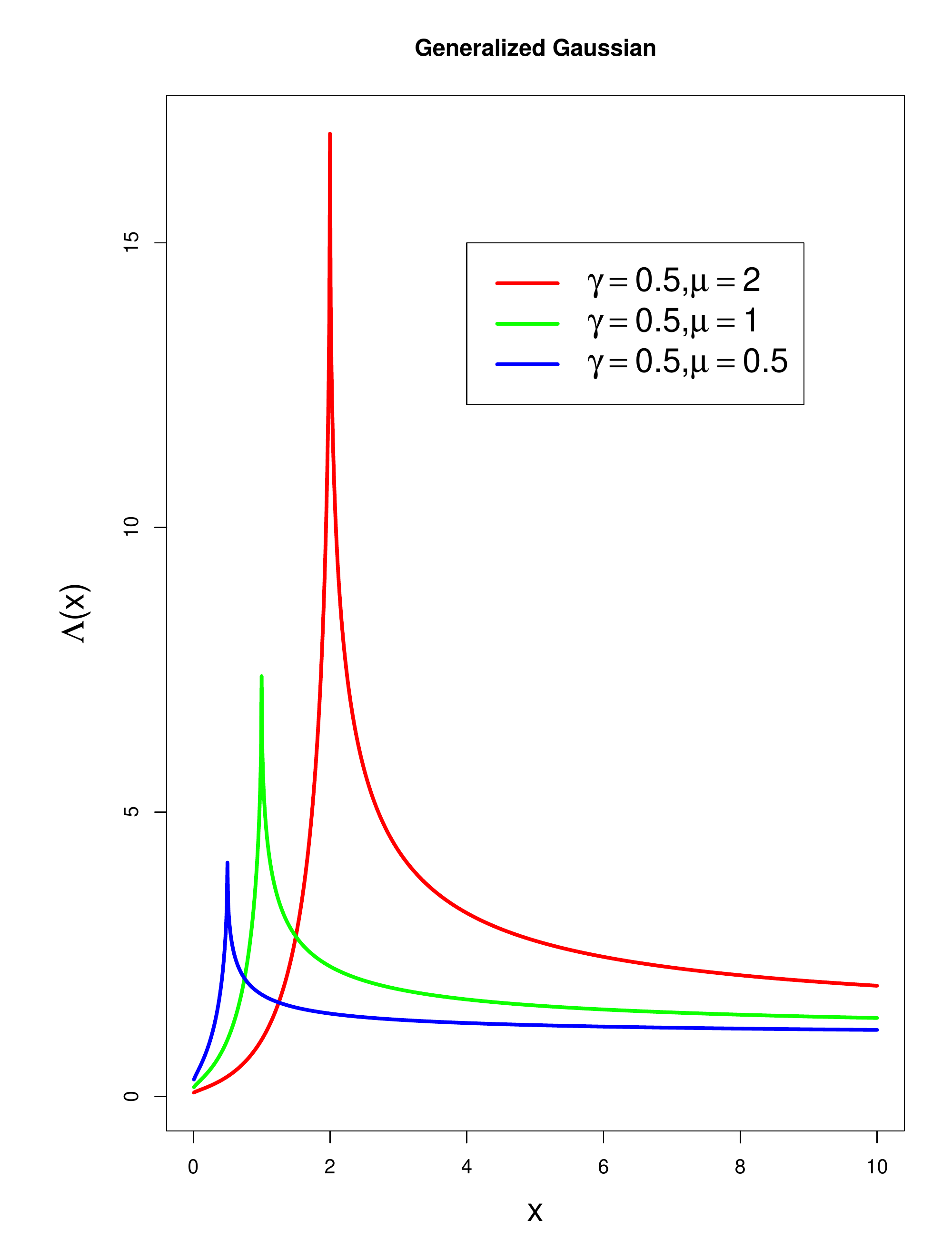}}
\subfigure[]{\label{fig:cauchy}\includegraphics[width=50mm,height=50mm]{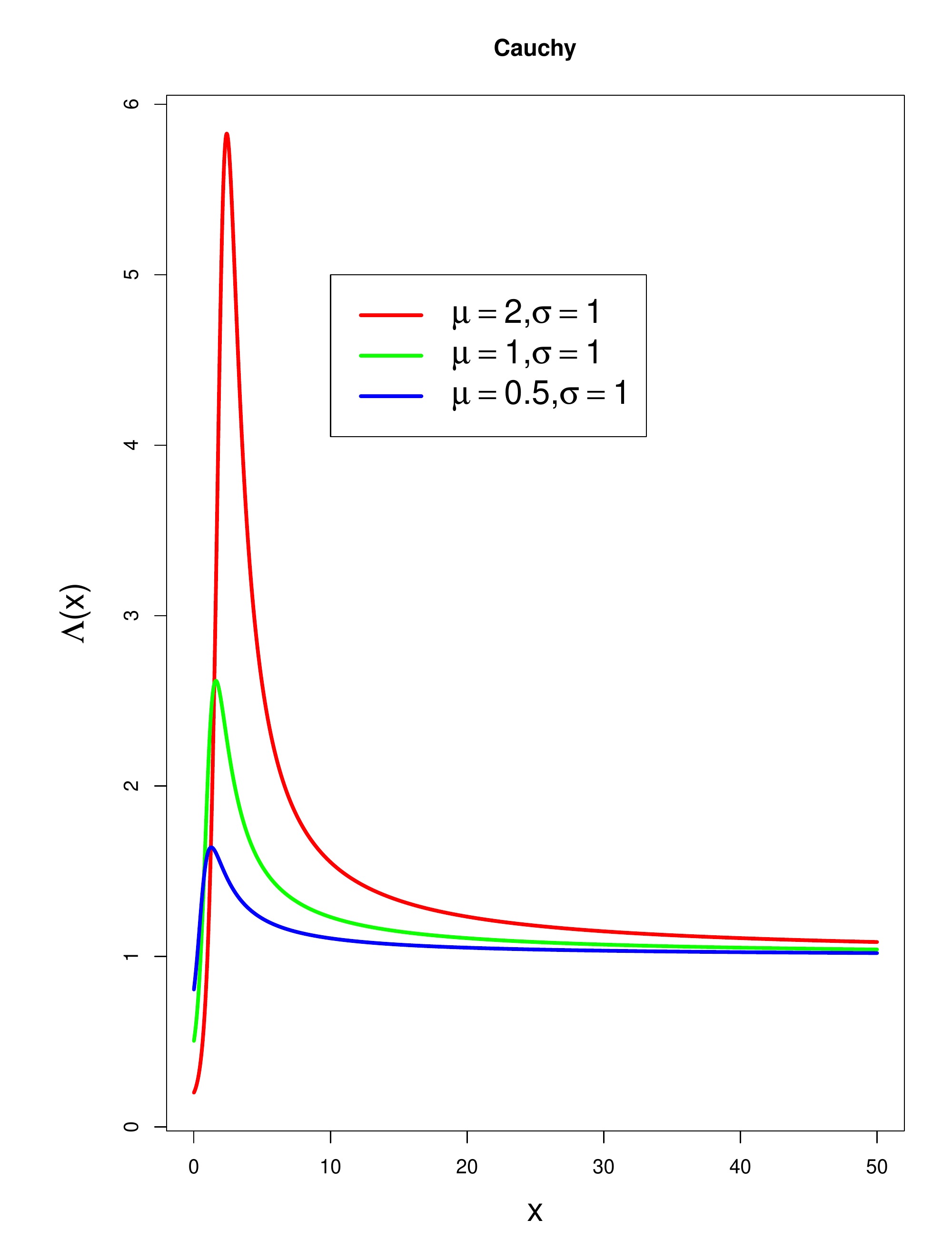}}         
\subfigure[]{\label{fig:lr:Golden}\includegraphics[width=50mm, height=50mm]{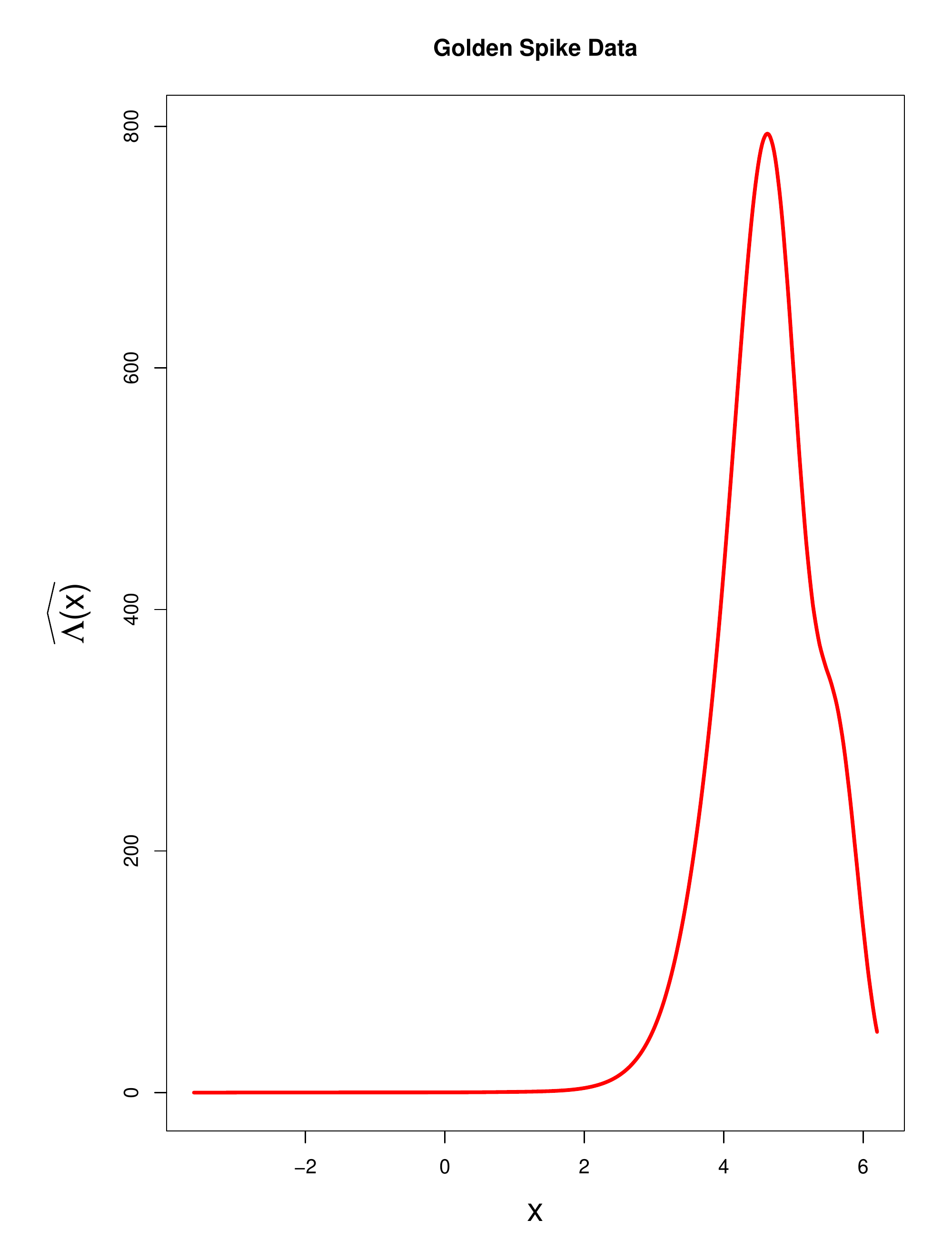}}
\caption{The likelihood ratios of different cases. The panels correspond to (a) the Gaussian case, (b) the generalized Gaussian case, (c) location-scale transformation of Cauchy distribution and (d) the estimated likelihood ratio of the Golden Spike data. 
}\label{fig:lr:model}
\end{figure}

The non-monotonicity of the likelihood ratio exists in both theories and applications. In Figure \ref{fig:lr:model}, we have plotted the likelihood ratio $\Lambda(x)$ for the following settings.
\begin{enumerate}
\item[(a)] Gaussian case: $f_0(x)=\phi(x)$ and $f_1(x)=\frac{1}{\sigma}\phi(\frac{x-\mu}{\sigma})$ where $\phi(x)$ is the density function of a standard normal random variable;
\item[(b)] Generalized-Gaussian case: $X|H_0\sim GN_\gamma(0)$ and $X|H_1\sim GN_\gamma(\mu), \mu>0$, where $GN_\gamma(\mu)$ is the family of generalized-Gaussian (Subbotin) distribution with density functions $\phi_{\gamma,\mu}(x)=C_{\gamma,\mu} e^{-\frac{|x-\mu|^\gamma}{\gamma}}$;

\item[(c)] Location-scale family: $f_0(x)$ be the probability density function of certain distributions, such as Cauchy, student's t-distributions, and $f_1(x)=\frac{1}{\sigma}f_0(\frac{x-\mu}{\sigma})$ be the location-scale transformation of $f_0(x)$;
\item[(d)] Estimated likelihood ratio from the Golden Spike data set from \citet{Choe:Bouttros:Michelson:Chruch:Halfon:2005}, which will be revisited in Section \ref{sec:data}.
\end{enumerate}

The decision in (\ref{EqThreshold04}) at first seems counter-intuitive because of the commonly-held assumption that extremely large statistics usually indicates strong evidence against the null based on the assumption of the MLR. However, such a convenient assumption does not hold in general for reasons, such as model mis-specification, heterogeneity, the existence of hidden variables and many others. In this article, we develop a method that agrees with the traditional method when MLR holds and is more accurate when the MLR condition does not hold.

Note that the computation of appropriate p-values under (5) is not obvious, since the event "the test statistics is at least as extreme as $X_i$" is not precisely defined. One could argue to use the likelihood ratio function to define "extremeness". We will discuss issues relating to this approach in the next section.

\subsection{Likelihood Ratio Test}\label{sec:intro:locfdr}
The famous Neyman-Pearson lemma, introduced in \cite{Neyman:Pearson:1928a, Neyman:Pearson:1928b, Neyman:Pearson:1933}, offers the most powerful test. The Neyman-Pearson lemma has a Bayesian interpretation.



\begin{exam}\label{example:1}
Consider a Bayesian classification problem where the goal is to classify  $X_i$'s, $i=1,2,\cdots,n$, into two groups, one consists of data generated from $U(0,1)$ and the other consists of data generated from the following distribution:
\begin{eqnarray*}
f_1(x)=\left\{ \begin{array}{cc} 
\frac{1}{l^2}n^{2\alpha}x, & \textrm{ if $x \le ln^{-\alpha}$,}\\
-\frac{1}{l^2}n^{2\alpha}(x- 2ln^{-\alpha}), & \textrm{ if $ln^{-\alpha}< x\le 2ln^{-\alpha}$,} \\
0, &\textrm{ if $2ln^{-\alpha}\le x\le 1$,} \end{array}\right.
\end{eqnarray*}
where $0<\alpha<1$, $l\le \frac{1}{2}n^\alpha$ are parameters.
\end{exam}

The ``likelihood''  that $X_i$  is from the first group  can be measured by the following posterior probability,
\begin{eqnarray*}
P\{\mbox{$X_i$ is from $U(0,1)$} | \boldsymbol{X} = \boldsymbol{x} \} = \frac{ (1-\pi_1)f_0(x_i) }{(1 - \pi_1)  f_0(x_i) +   \pi_1 f_1(x_i)}   = \frac{1-\pi_1}{(1-\pi_1)  + \pi_1 f_1(x_i)}.  
\end{eqnarray*}
This is also called the local fdr, denoted as $fdr_i(\vx)$ (\cite{Efron:Tibshirani:Storey:Tusher:2001}, \cite{Efron:2008, Efron:2010b}, \cite{Sun:Cai:2007}, \cite{Cao:Sun:Kosorok:2013}, \cite{He:Sarkar:Zhao:2015}, \cite{Liu:Sarkar:Zhao:2016}).
The Bayesian classification rule would simply classify $X_i$ into the first group if and only if: 
\begin{equation}\label{eqn:bayes}
fdr_i(\boldsymbol{x})   \geq \frac{1}{2},
\end{equation}
which agrees with the procedure defined in (\ref{EqThreshold04}) when $T$ and $S$ are chosen appropriately.

When assuming the two-group model (\ref{model}), then the local fdr is
\begin{equation}\label{def:locfdr}
fdr_i(\vx) = P(H_{0i}|\vx ) = \frac{ (1-\pi_1) f_0(x_i)}{ f(x_i)}.
\end{equation}
The local fdr based approach originates from the Bayesian classification rule. It is shown that the decision $\delta_i=1(fdr_i(\vx)\le c)$ for some appropriately chosen $c$ is optimal (\cite{Sun:Cai:2007, He:Sarkar:Zhao:2015}). However, the local fdrs rely on the probability density function $f(x)$. There have been many attempts, including \cite{Efron:Tibshirani:Storey:Tusher:2001}, \cite{Efron:2008}, \cite{Sun:Cai:2007}, \cite{Sun:Cai:2009}, and \cite{Cao:Sun:Kosorok:2013}, to derive a data-driven version of it by estimating these local fdrs. However, developing a good non-parametric estimator of the probability density function is a known challenging problem.

\begin{figure}
\centering
\includegraphics[width=75mm,height=75mm]{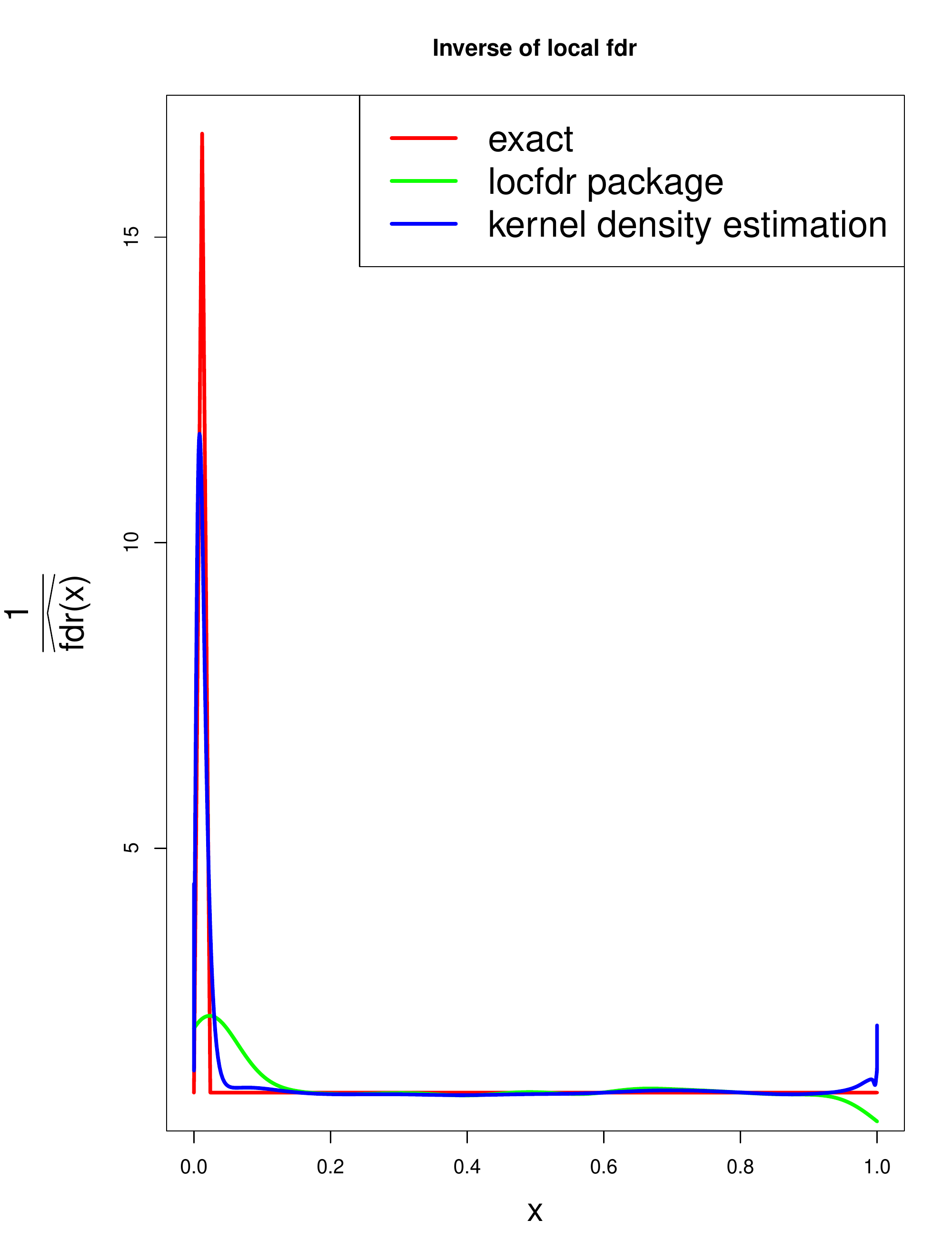}
\caption{Plot of the $(fdr(x))^{-1}$ and its estimate. The red solid line is the inverse of local fdrs based on the true probability density function. The green curve corresponds to the inverse of estimated local fdrs based on the {\it locfdr} package. The blue curve corresponds to the inverse of estimated local fdrs based on the kernel density estimator.
}\label{fig:locfdr:density}
\end{figure}

In Example \ref{example:1}, set $n=5,000$, $l=1.2$, $\alpha=0.5$, $\beta=0.2$, $\pi_1=n^{-\beta}=18.2\%$ and set $q$, a desired \mfdrspace level, as $0.1$. 
We then generate a random sample $X_i, i=1,2,\cdots, n$ and $Y_i = \Phi^{-1}(1-X_i)$ to be the transformed data. We calculate the local fdrs according to (\ref{def:locfdr}) where the marginal probability density function $f(x)$ is estimated using either {\it locfdr} package or the kernel density estimator.
In Figure \ref{fig:locfdr:density}, we plot the inverse of the true local fdr $ (fdr(x))^{-1}$ (red curve), the inverse of an estimated local fdr based on the {\it locfdr} package (green curve), and the inverse of an estimated local fdr based on the kernel density estimator (blue curve). Both methods, which estimate the probability density function of the test statistic, smooth the area around the spike and fail to capture the spike around 0. The estimation based on the {\it locfdr} performs poorly in this case as it completely missed the spike of the mixture distribution.

After estimating the local fdrs, these quantities are ordered increasingly as
\[
fdr_{(1)}(\vx)\le fdr_{(2)}(\vx)\le \cdots \le fdr_{(n)}(\vx).
\]
Additionally, let
\[
R = \max_r\left\{r: \frac{1}{r}\sum_{i=1}^r fdr_{(i)}(\vx) \le q \right\}.
\]
We reject those hypotheses corresponding to the first $R$ smallest local fdrs (\cite{Sun:Cai:2007, Sarkar:Zhou:Ghosh:2008}). We replicate these steps 100 times to calculate the average number of true rejections (ET), average number of false rejections (EV), and \mfdr. For comparison, the results of BH method with the p-values given as $p_i=P(U\le X_i)=X_i$ and the proposed method (CLAT) are also reported in Table \ref{tab:locfdr:comp}. Due to the difficulty of estimating the probability density function, the data-driven version of local fdr based approaches don't provide reliable decision for this example. BH method also fails because $\Lambda(x)$ is not monotone on the left side. The CLAT works well and rejects the highest number of hypothesis subject to control of the \mfdrspace at $q$-level.


\begin{table}
  \centering
  \begin{tabular}{|c|c|c|c|}
    \hline
    & ET & EV & \mfdr\\
    \hline 
    locfdr package based method & 0.71 & 0.20 & 0.07\\
    \hline
    Kernel density estimation based method & 0.22 & 0.14 & 0.05\\
    \hline
    BH method & 0  & 0.05 & 0.05\\
    \hline
    CLAT(Proposed method) & 390 & 37 & 0.07 \\
    \hline
  \end{tabular}
  \caption{This table summarizes the average number of true rejections, average number of false rejections, and the \mfdrspace of various methods when setting $q$ as $0.1$. It is seen that all the methods control the error rates at a desired level. The CLAT rejects a good number of hypotheses. But all the other methods reject very small number of hypotheses on average.
  }\label{tab:locfdr:comp}
\end{table}

In Section \ref{sec:pvalue}, we consider the commonly used p-value that relies on the null distribution only. One reviewer suggested other forms of the p-value that depend on the likelihood ratio $\Lambda(x)$. Such a p-value could lead to an optimal testing method in theory; however, like many local fdr based methods, a data-driven version would depend on an estimation of the probability density function. Thus, it faces the same difficulty as local fdr based methods.

\subsection{CLAT}\label{sec:comprise}
In Section \ref{sec:pvalue}, we show that traditional p-value based approaches are not optimal for cases with non-monotone likelihood ratio (Non-MLR). In Section \ref{sec:intro:locfdr}, it is shown that local fdr based approaches are optimal, but rely on an estimation of the probability density function. In this section, we introduce a new method that is optimal for many Non-MLR cases and relies on the estimation of the cumulative distribution function only.

Motivated by (\ref{EqThreshold04}), we consider the following rejection interval $\mathbbm{I}_F(q)$,
\begin{equation}\label{eq:gen:bh}
\mathbbm{I}_F(q)   =  argmax_{\mathbbm{I}_{\{t,s\}}}  \biggl\{ \int_{\mathbbm{I}_{\{t,s\}}}   dF:  q\int_{\mathbbm{I}_{\{t,s\}}}   dF      \ge      \int_{\mathbbm{I}_{\{t,s\}}}(1-\pi_1)   dF_0 \biggr\}.
\end{equation}
A hypothesis is rejected if the test statistic falls into this interval. It can be shown that $\mathbbm{I}_F(q)$ offers an rejection interval which is optimal among all rejection intervals.
To derive a data driven version of $\mathbbm{I}_F(q)$, one can replace the cumulative distribution function $F(x)$ with the empirical distribution function $F_n(x)=\frac{1}{n}\sum_{i=1}^n\mathbbm{1}(X_i\le x)$ and replace $\pi_1$ by an appropriate estimator (see Remark 1 below). The data-driven rejection interval $\mathbbm{I}_n(q)$ is thus written as
\begin{equation}\label{intro:eq:adj:bh}
  \mathbbm{I}_{F_n}(q)   =  argmax_{\mathbbm{I}_{\{t,s\}}}  \biggl\{ \int_{\mathbbm{I}_{\{t,s\}}}   dF_n:  q\int_{\mathbbm{I}_{\{t,s\}}}   dF_n      \ge     \int_{\mathbbm{I}_{\{t,s\}}} (1-\pi_1)  dF_0 \biggr\}.
\end{equation}

Unlike many local fdr based approaches which require an estimation of the probability density function, this method relies on the empirical distribution function, which is free from choosing tuning parameters and, according to the well-known DKW theorem (\cite{Dvoretzky:Kiefer:Wolfowitz:1956}), converges to the true cumulative distribution function uniformly at a fast rate. 
This new method yields good theoretical properties and methodological performance. It successfully combines the advantages of both p-value and local fdr based approaches. We call this method ``{\bf C}df and {\bf L}ocal fdr {\bf A}ssisted multiple {\bf T}esting method (CLAT)''.

\subsection{Algorithm}\label{sec:algorithm}
There is an issue when implementing method (\ref{intro:eq:adj:bh}). The estimation error of the empirical distribution function has an order of $O(\frac{1}{\sqrt{n}})$. When the signal is sparse, the estimation error could dominate the probability of the rejection region. The proportion of data that falls in an erroneous interval surrounding zero with length of $O(\frac{1}{\sqrt{n}})$ could be even larger than that of the ideal interval $\mathbbm{I}_F(q)$. To avoid this, we restrict the length of $\mathbbm{I}_{F_n}(q)$ such that $Length(\mathbbm{I}_{F_n}(q))\ge \frac{C\log{n}}{\sqrt{n}}$. The choice of the constant $C$ is not critical and is chosen as 2 in the following algorithm.

\begin{algorithm}[H]
	\caption{CLAT. \label{alg:gbh}}
	\begin{algorithmic}[1]
		\vspace*{1mm}
		\\
		Calculate the p-values $p_i$ for each hypothesis as $p_i=1-F_0(x_i)$;\\
		\vspace*{1mm}
		Order the p-values increasingly as $p_{(1)}\le p_{(2)}\le \cdots\le p_{(n)}$;
		\\
		\vspace*{1mm}
		Find $I$ and $J$ such that $J-I=M$ where
		\[
		M=\max\left\{j-i: i\le j, p_{(j)}-p_{(i)} \le \frac{q}{1-\pi_1} \frac{j-i}{n}, |F_0^{-1}(p_{(j)})-F_0^{-1}(p_{(i)})|>\frac{2\log{n}}{\sqrt{n}}\right\};
		\]
		\\
		\vspace*{1mm}
		(a) If $J>I$, then reject the $i$-th hypothesis $H_i$ where $ p_{(I)}\le p_i\le p_{(J)}$ and accept the rest; 
		
		(b) If $J=I$, accept all the hypotheses.

	\end{algorithmic}
\end{algorithm}

\begin{remark}
	In Algorithm \ref{alg:gbh}, we assume $\pi_1$ is known. If it is not, one could either replace it with a reliable estimator $\hat{\pi}_1$ or set $\pi_1$ as zero and the resultant method still controls the \mfdrspace at $q$-level.
  \end{remark}

In Step 3 of Algorithm \ref{alg:gbh}, the computational complexity of direct searching $I$ and $J$ is $O(n^2)$, which is not feasible when the number of hypotheses is large. We substitute it with the following novel algorithm.

Note that the key constraint is $p_{(j)}-p_{(i)} \le \frac{q}{1-\pi_1} \frac{j-i}{n}$ which can be rewritten as
\begin{equation}\label{trans:restrict}
\frac{qi}{n(1-\pi_1)} - p_{(i)} \le \frac{qj}{n(1-\pi_1)} - p_{(j)}.
\end{equation}
Let $T_i= \frac{qi}{n(1-\pi_1)}-p_{(i)}, (1\le i\le n)$ and order $T_i$ increasingly as $ T_{(1)}\le T_{(2)}\le \cdots \le T_{(n)}$. Let $l_i$ be the index such that $T_{l_i}= T_{(i)}$.
Then for any two integers $l_i,l_j$ with $i<j$, 
\[
p_{(l_j)}-p_{(l_i)} \le q\frac{l_j-l_i}{n(1-\pi_1)} .
\]
This problem can be simplified as finding the maximum value of $l_j-l_i$ where $i<j$. For each $j$, we only need to calculate the difference between $l_j$ and $\min_{1\le k\le j}l_i$, thereby requiring us to scan the whole sequence $l_i$'s once. 

Based on this, we replace Step 3 of Algorithm \ref{alg:gbh} by the following:
\begin{algorithm}[H]
			\caption{Step 3'. }
	\begin{algorithmic}[1]
		\vspace*{1mm}
		\\
		Calculate $T_i=\frac{qi}{n}-p_{(i)}$, order $T_i$'s increasingly and obtain $l_i$;
		\\
		Let $I = 1, J =1, i_{temp}=l_1$ and $MAXDIFF=0$. 
		
		For $j$ in $1:n$,
		
		(a) If $l_j<i_{temp}$, let $i_{temp}=l_j$;
		
		(b) If $T_{l_j}\ge 0$ and $l_j>MAXDIFF$, let $I=1, J=l_j$, and $MAXDIFF=l_j$;
		
		(c) If $l_j-i_{temp}>MAXDIFF$ and $|F_0^{-1}(p_{(l_j)})-F_0^{-1}(p_{(i_{temp})})|>\frac{2\log{n}}{\sqrt{n}}$, let $J = l_j$, $I =i_{temp}$, and $MAXDIFF= l_j-i_{temp}$;
		
	\end{algorithmic}
\end{algorithm}

\begin{remark}
  Algorithm \ref{alg:gbh} is designed for the right-sided test. For the left sided test, we calculate p-values as $p_i=F_0(x_i)$ and then follow Steps 2, 3' and 4. When testing two sided hypotheses, we apply the algorithm to the left-sided and right-sided p-values at level $q$ respectively to get two rejection sets $\mathbbm{I}_{F_n}^+(q)$ and $\mathbbm{I}_{F_n}^-(q)$. The final rejection set is the union of these two. Namely, $\mathbbm{I}_{F_n}(q) = \mathbbm{I}_{F_n}^+(q) \cup \mathbbm{I}_{F_n}^-(q)$.
\end{remark}

The remaining part of the paper is organized as follows. In Section \ref{sec:main}, we introduce the oracle and data-driven version of the CLAT and study their theoretical properties. Sections \ref{sec:simulation} and \ref{sec:data} include simulations and data analysis, all of which show that CLAT is powerful subject to control of the error rate. We provide technical proofs in Section \ref{sec:app}.

\section{Main Result}\label{sec:main}
\subsection{Oracle Procedure}\label{sec:ideal}

To save space and simplify the argument, we focus on the right-sided test. Similar results can be obtained for the left-sided and two-sided test with appropriate adjustment. Assume that the test statistic $X_i$'s are continuous random variables with support of $(-\infty,\infty)$ and both $\pi_1$ and $f_1(x)$ $(F_1(x))$ are known. We start with the discussion of an oracle version of BH method (\cite{Benjamini:Hochberg:1995}) where a hypothesis is rejected if the corresponding test statistic is greater than or equal to $T^*_{BH}(q)$, defined as
\[
T^*_{BH}(q)=argmin_t\left\{ q\int_t^{+\infty}dF(x)\ge \int_t^{+\infty}dF_0(x) \right\}.
\]
This interval does not depend on the non-null proportion $\pi_1$, and is called distribution-free (\cite{Genovese:Wasserman:2002}).
If there exists reliable information of $\pi_1$, one can choose a less conservative $T_{BH}(q)$ as 
\[
T_{BH}(q)=argmin_t\left\{ q\int_t^{+\infty}dF(x)\ge \int_t^{+\infty}(1-\pi_1) dF_0(x) \right\}.
\]
Let $\mathbbm{I}_{BH}(q)=[ T_{BH}(q), \infty)$, which is referred to as the oracle BH interval. Note that $\mathbbm{I}_F(q)$ reduces to the BH interval when setting $s=+\infty$.\\

When the likelihood ratio is not monotone, $\mathbbm{I}_{BH}(q)$ does not lead to the optimal rejection interval. 
To observe this, consider the following example where 
\begin{equation}\label{equ:toy:example}
f(x)=(1-\pi_1)\phi(x) + \pi_1\frac{1}{\sigma}\phi(\frac{x-\mu}{\sigma}).
\end{equation}
 Let $n=100,000$ and $\pi_1=n^{-\beta}$ be the proportion of non-null hypotheses. For different choices of $(\beta,\mu,\sigma)$, we randomly generate an independent sequence $X_1,X_2,\cdots,X_n$ according to (\ref{equ:toy:example}) and then order them decreasingly as $X_{(1)}\ge X_{(2)}\ge\cdots\ge X_{(n)}$. 
Let $r$ be the smallest $k$ such that $X_{(k)}$ is generated from the alternative distribution. Namely,
\[
r=\min_k\{k: X_{(k)} \textrm{ is generated from the alternative distribution} \}.
\]
We replicate this step 100 times, calculate the average $r$, and report this number in the fourth column of Table \ref{tab:BH}. For instance, when $\beta=0.6$, $\sigma=0.8$ and $\mu=1.5$, the average $r$ is 47.1, implying that, on average, the first 46 largest observations are generated from the null hypothesis. Hence, the interval $\mathbbm{I}_{BH}(q)$ does not provide a good choice of a rejection interval.

\begin{table}
	\centering
	\begin{tabular}{|c|c|c|c|}
		\hline
		$\beta$&$\sigma$&$\mu$&Ave $r$\\
		\hline
		0.7&0.8&2.0&38.77\\
		\hline
		0.7&0.5&2.5&32.5\\
		\hline
		0.6&0.8&1.5&47.12\\
		\hline
	\end{tabular}
	\caption{ Assume that $f(x)=(1-\pi_1)\phi(x)+ \pi_1\frac{1}{\sigma}\phi(\frac{x-\mu}{\sigma})$ where $\pi_1=p^{-\beta}$ and $p=100,000$.
		This table summarizes the average $r$ for each parameters setting.
	}\label{tab:BH}
\end{table} 


  
This example motivates us to select an oracle rejection set $\mathbbm{S}_F(q)$ as
\begin{equation}\label{eq:ideal:region}
\mathbbm{S}_F(q)   =  argmax_{\mathbbm{S}\subset R^1}  \biggl\{ \int_{\mathbbm{S}}   dF: q \int_{\mathbbm{S}}   dF      \ge   \int_{\mathbbm{S}}   (1-\pi_1)dF_0 \biggr\}.
\end{equation}
Note that the decision based on this rejection set maximizes the probability of rejections subject to control of \mfdrspace at $q$-level. It can be shown that \mfnrspace is minimized and this decision is optimal.
According to \cite{Sun:Cai:2007} and \cite{He:Sarkar:Zhao:2015}, among all the sets which controls the \mfdrspace at a given $q$ level, the one maximizing the average power is the set $\{x: fdr(x)\le c\}$ for a properly chosen constant $c$. Note that $fdr(x)$ is decreasing with respect to the likelihood ratio $\Lambda(x)$. The oracle rejection set $\mathbbm{S}_F(q)$ can also be chosen as the set of $x$ such that $\Lambda(x)$ exceeds a certain level. We thus offer the following theorem.

\begin{theorem}\label{thm:1}
  \begin{enumerate}
  	\item[(a)] When $\Lambda(x)$ is monotone increasing, then the oracle rejection set $\mathbbm{S}_F(q)$, the oracle interval $\mathbbm{I}_F(q)$, and the oracle BH interval $\mathbbm{I}_{BH}(q)$ are the same;
  	\item[(b)] When $\mathbbm{S}_F(q)$ is a finite interval $[t,s]$, then $\mathbbm{I}_F(q)$ agrees with $\mathbbm{S}_F(q)$ and is optimal; however, the $\mathbbm{I}_{BH}(q)$ is not optimal.
  \end{enumerate}
\end{theorem}

Existing literature discusses how to find the rejection set with several discussions aimed at exploring whether such a set exists (\cite{Zhang:Fan:Yu:2011}). Next theorem gives a necessary condition of the existence of a non-empty rejection set.

\begin{theorem}\label{thm:prop:ness}
	Assume the two-group model (\ref{model}). If $\max_x \Lambda(x)< q'$ where $q'=\frac{(1-q)(1-\pi_1)}{q\pi_1}$, then 
for any set $\mathbbm{U}=\cup_{i=1}^\infty \mathbbm{I}_i$ where $\mathbbm{I}_i$ are disjoint intervals,
\[
(1-\pi_1)\int_{\mathbbm{U}}dF_0(x)dx> q\int_{\mathbbm{U}}dF(x).
\]
If we reject a hypothesis when the test statistics falls in $U$, then 
\[
\mfdr = \frac{(1-\pi_1)\int_{\mathbbm{U}}dF_0(x)dx }{ \int_{\mathbbm{U}}dF(x) } >q.
\]
\end{theorem}


When $\Lambda(x)$ is monotone increasing, intuitively, one would conjecture that \mfdrspace can be arbitrarily small as $T_{BH}(q)$ in $\mathbbm{I}_{BH}(q)$ moves toward infinity. Unfortunately, this intuition is no longer true. 
One counter-example is the case when $f_0$ and $f_1$ are the density function of a student's t-distribution and non-central student's t-distribution with $d$ degree of freedom. The likelihood ratio is monotone increasing with an upper limit. Consequently, there is a lower limit of the \mfdrspace level that one can control. When setting the desired \mfdrspace level to be smaller than this limit,
there is no rejection set $\mathbbm{S}$ such that $\int_\mathbbm{S} dF(x)>0$ and \mfdrspace is less than or equal to the desired level based on this rejection set.


On the other hand, if $\max_x \Lambda(x)>q'$, then under certain regularity conditions, such a rejection set exists. 
\begin{theorem}\label{thm:prop:suff}
Assume that $\max_x \Lambda(x)>q'$. Let $c_1$ and $c_2$ be the solutions of $\Lambda(x)=q'$.
Assume 
that $\Lambda(x)>q'$ for all $x\in (c_1,c_2)$. Then the \mfdrspace based on the rejection interval $[c_1,c_2]$ is less than or equal to $q$.

\end{theorem}

\begin{theorem}\label{thm:prop:suff2}
  If $\Lambda(x)$ is monotone and $\max_x \Lambda(x)>q'$. Let $c$ be a constant such that $\Lambda(c)=q'$. Then $(c',\infty)\subset \mathbbm{I}_{F}(q)$.
\end{theorem}

In theory, the rejection set $\mathbbm{S}_F(q)$ comprise the union of multiple disjoint intervals. But this rarely happens in practice. We therefore focus on the rejection interval $\mathbbm{I}_F(q)$ for the one-sided test. For the two-sided test, the rejection set is chosen as $\mathbbm{I}_F^+(q)\cup \mathbbm{I}_F^-(q)$ where $\mathbbm{I}_F^+(q)$ and $\mathbbm{I}_F^-(q)$ are the rejection interval based on the right-sided test and left-sided test respectively.

\subsection{Convergence rate of the generalized BH procedure}\label{sec:asym}
In Section \ref{sec:ideal}, we discussed the oracle interval $\mathbbm{I}_F(q)$ assuming $F(x)$ is known. When it is unknown, we can estimate it by the empirical distribution function and
obtain the data-driven version of $\mathbbm{I}_F(q)$.
DKW's inequality guarantees that $P(\sup_x|F_n(x)-F(x)|>\epsilon)\le 2e^{-2n\epsilon^2}$.
Therefore, we expect that the empirical interval could mimic the oracle interval well. 

Before stating the theorem, we introduce some notations. 
Let $s(a,b)=(1-\pi_1)\int_a^bdF_0-q\int_a^bdF$, $s_n(a,b)=(1-\pi_1)\int_a^bdF_0-q\int_a^bdF_n$. Note that $s(a,b)\le 0$ imples that the \mfdrspace based on the rejection interval $[a,b]$ is less than or equal to $q$. Let $c_1, c_2$ and $q'$ be the constants defined in Theorems \ref{thm:prop:ness} and \ref{thm:prop:suff}. Let $b_a(F)=argmax_b \{b: s(a,b) \le 0\}$. Then $[a, b_a(F)]$ is the longest rejection interval starting from $a$ which controls \mfdrspace at $q$-level. Let $g(a)= F(b_a(F))-F(a)$ be the probability of the rejection set $(a, b_a(F))$. Similarly, define $b_a(F_n)=argmax_b \{b: s_n(a,b) \le 0\}$ as the empirical version of $b_a(F)$ and $g_n(a)=F_n(b_a(F_n))- F_n(a)$ be the proportion of hypotheses being rejected.

\begin{theorem}\label{thm:conRate}
Assume that $f_0, f_1, h \in C^1(R)$ and conditions in Theorem \ref{thm:prop:suff} hold and $q'\int_{-\infty}^{c_2}f_0(x)>q$.
Let $[a_0, b_{a_0}(F)]$ be the ideal rejection interval.
Assume that $g_n(a)$ attains the maximum at $a=a_n$ and a hypothesis is rejected if the test statistic falls between $a_n$ and $b_{a_n}(F_n)$.
Then $\limsup_{n\to\infty} \mfdr\le q$ and there exists a constant $C$ such that
\begin{equation}
P(| g_n(a_n)-g(a_0)|>C\epsilon) \le 2e^{-2n\epsilon^2}.
\end{equation}
\end{theorem}

{\it Remark: According to this theorem, the CLAT controls the \mfdrspace at $q$-level asymptotically and the proportion of hypotheses being rejected converges to the probability of the ideal  rejection interval $\mathbbm{I}_F(q)$ with a rate of $O_P(\frac{1}{\sqrt{n}})$. 
}



\section{Simulation}\label{sec:simulation}
In this section, we use simulations to compare various approaches, namely, CLAT, BH method, and local fdr based methods. For the other local fdr based methods, the probability density function of the test statistic  are estimated using (i) {\it locfdr} package using , (ii) kernel density estimation (SC method \cite{Sun:Cai:2007}), and (iii) EM algorithm. The steps of EM algorithms are outlined in the supplementary materials. These three methods are denoted as Lfdr-locfdr, Lfdr-SC, Lfdr-EM. 

For the following three cases, assume that the test statistic $X_i$ are generated from the distribution $X_i\iid (1-\pi_1)f_0(x)+\pi_1 f_1(x)$ for $i=1,2,\cdots, n$ where $\pi_1=n^{-\beta}$. We also include the oracle method when assuming all the parameters are known. This method is denoted as Lfdr-oracle. 

{\bf Case I:} $f_0(x)=\phi(x)$, the density function of a standard normal distribution and 
\[
f_1(x)\sim p_1\pi_1 \frac{1}{\sigma}\phi(\frac{x-\mu}{\sigma}) + (1-p_1)\pi_1 \frac{1}{\sigma}\phi(\frac{x+\mu}{\sigma}).
\]
The parameters are $\beta, p_1, \mu$, and $\sigma$.

{\bf Case II:} $f_0(x)=t_d(x)$ where $t_d(x)$ the density function of student's t-distribution with degrees of freedom $d$. The $f_1(x)$ is a mixture of two location-scale transformation of t-distribution, namely,
\[
f_1(x)\sim p_1\pi_1 \frac{1}{\sigma}t_d(\frac{x-\mu}{\sigma}) + (1-p_1)\pi_1 \frac{1}{\sigma}t_d(\frac{x+\mu}{\sigma}).
\]
The parameters are $d, \beta, p_1, \mu$, and $\sigma$.

{\bf Case III:} $f_0(x)=1(0\le x\le 1)$ be the density function of a uniform distribution. The $f_1(x)$ is given in Equation (\ref{equ:toy:example}) of Example \ref{example:1}.
The parameters are $\beta, \alpha$ and $\beta$.
\\

Our current theory is based on the independence assumption. In the following example, we run the simulation when the test statistic are dependent. \\

{\bf Case IV:} For given parameters $\alpha, \beta, p_1, \mu$ and $\sigma^2$, generate $X_i$'s according to Case I. Let $Z \sim N(0, \sigma_2^2)$. Let $Y_i = \frac{X_i+Z}{\sqrt{1+\sigma_2^2}}$. Then the correlation between $Y_i$ and $Y_j$ can be written as
\begin{eqnarray*}
	\rho(Y_i, Y_j) = \left\{ 
	\begin{array}{rl}
		\frac{\sigma_2^2}{1+\sigma_2^2}, & X_i\sim f_0(x), X_j\sim f_0(x);\\
		\frac{\sigma_2^2}{\sigma^2+\sigma_2^2}, & X_i\sim f_1(x), X_j\sim f_1(x); \\
		\frac{\sigma_2^2}{\sqrt{(1+\sigma_2^2)(\sigma^2+\sigma_2^2)}}, & X_i\sim f_0(x), X_j\sim f_1(x), or X_i\sim f_1(x), X_j\sim f_1(x).
		\end{array}
	\right.
\end{eqnarray*}

\begin{figure}
	\centering
	\includegraphics[width=50mm, height=50mm]{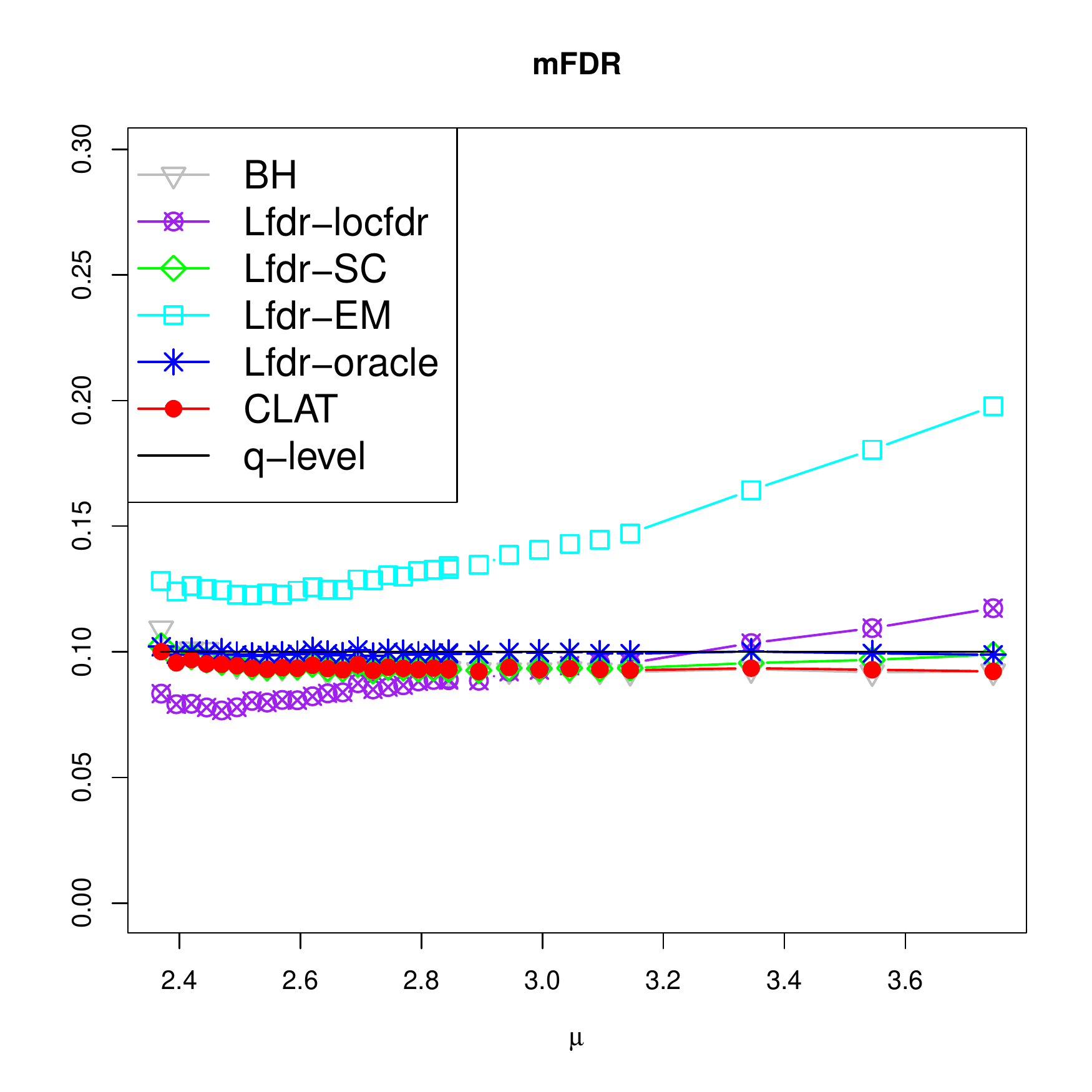}
	\includegraphics[width=50mm, height=50mm]{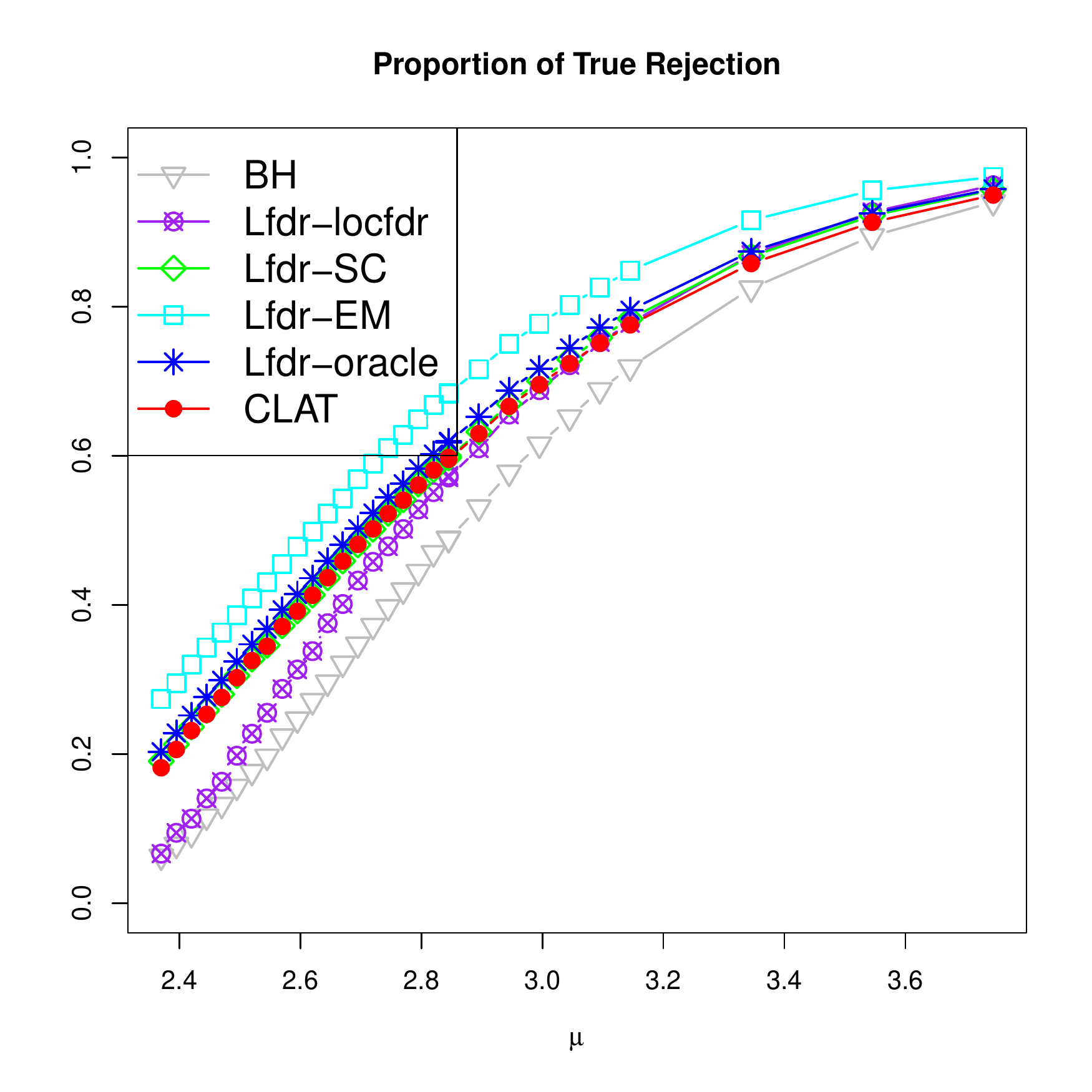}
	\caption{Case I: plot of the \mfdrspace and the average percentages of true rejections. The parameters are chosen as $n=5,000$, $\beta=0.3$, $p_1=0.9$, $\sigma=0.7$ and $\mu$ various from 2.4 to 3.8.
	}\label{fig:normal:1}
\end{figure}

\begin{figure}
	\centering
	\includegraphics[width=50mm, height=50mm]{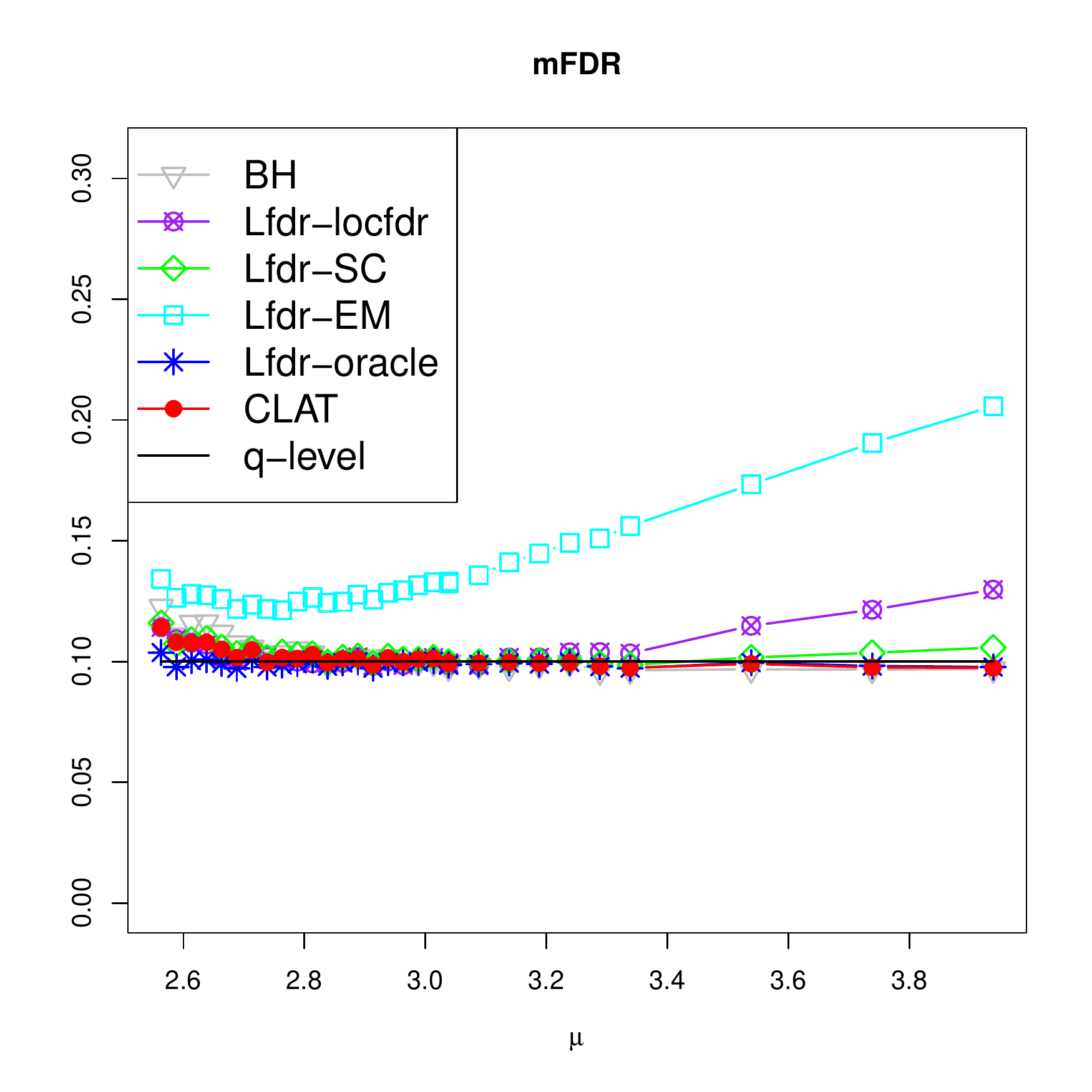}
	\includegraphics[width=50mm, height=50mm]{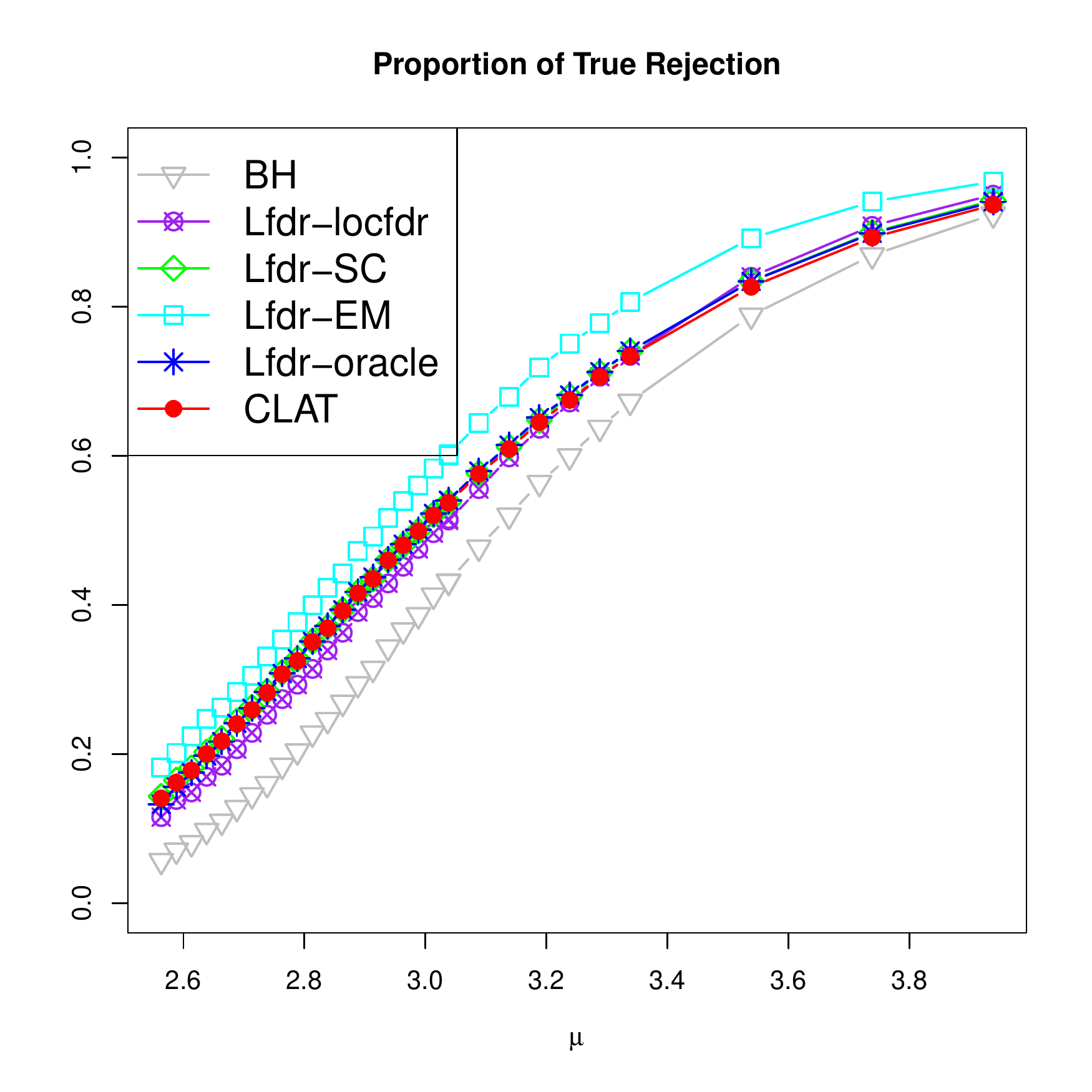}
	\caption{Case I: plot of the \mfdrspace and the average percentages of true rejections. The parameters are chosen as $n=5,000$, $\beta=0.4$, $p_1=0.9$, $\sigma=0.7$ and $\mu$ various from 2.6 to 4.0.
	}\label{fig:normal:2}
\end{figure}

\begin{figure}
	\centering
	\includegraphics[width=50mm, height=50mm]{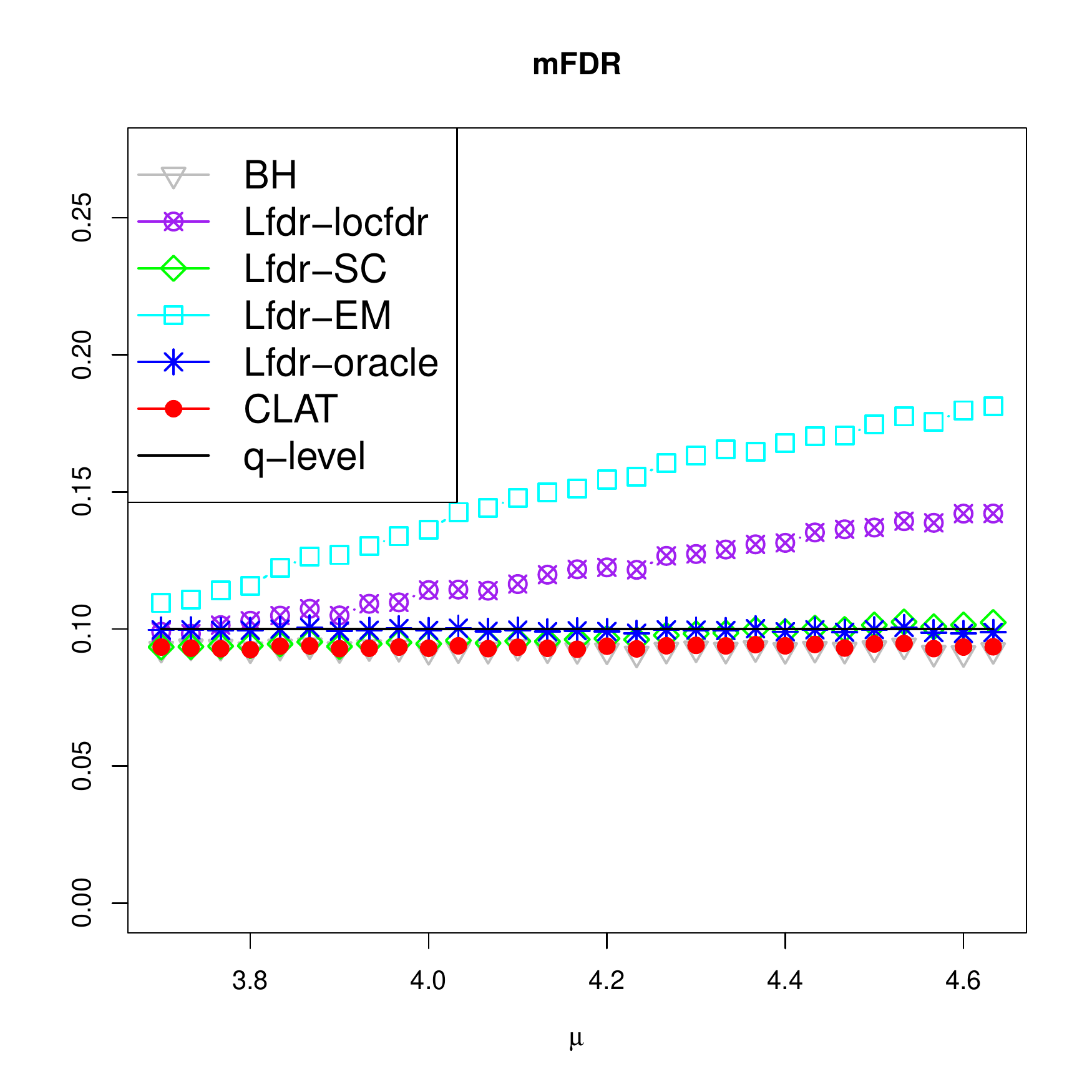}
	\includegraphics[width=50mm, height=50mm]{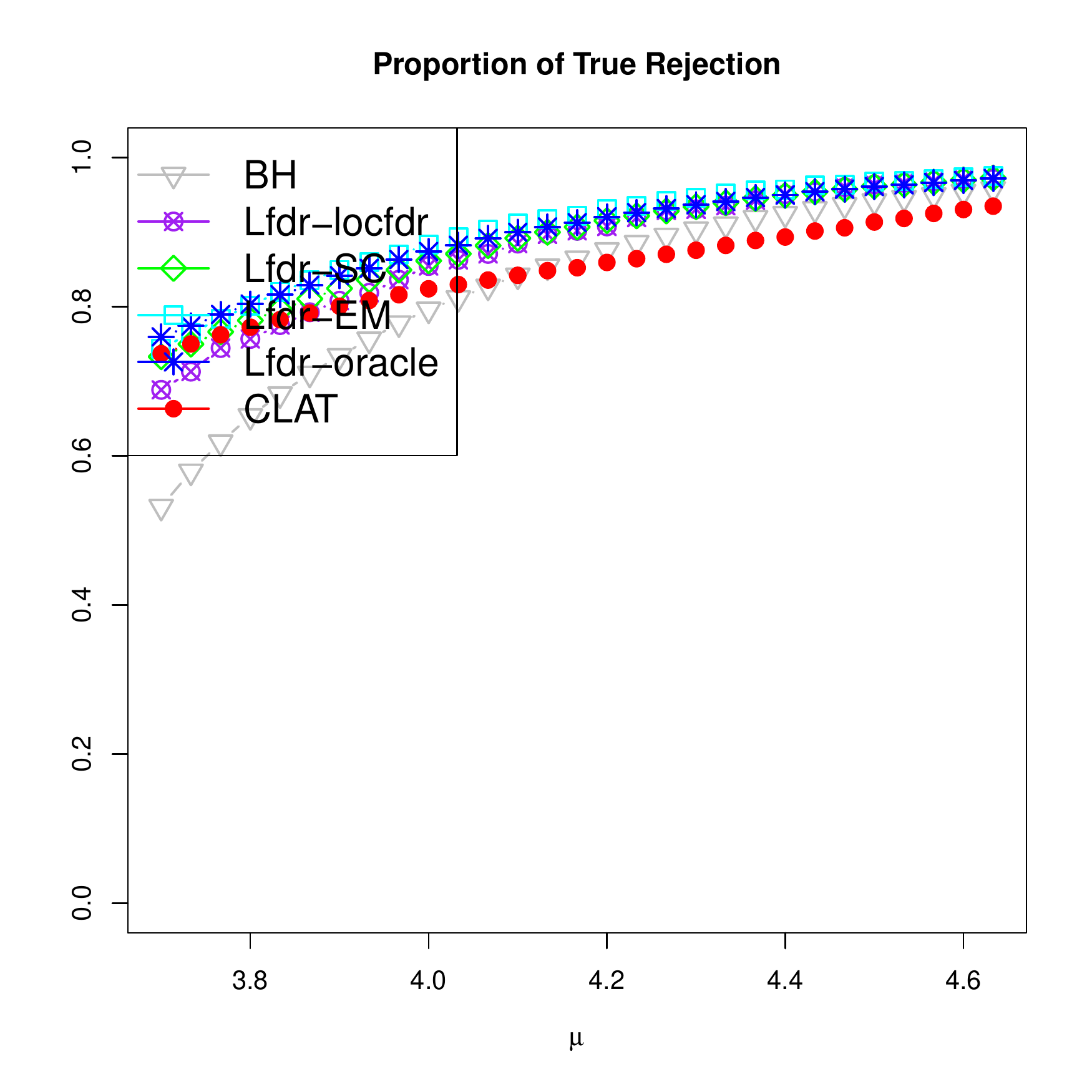}
	\caption{Case II: plot of the \mfdrspace and the average percentages of true rejections. The parameters are chosen as $n=5,000$, $df=10$, $\beta=0.3$, $p_1=0.9$, $\sigma=0.7$ and $\mu$ various from 3.1 to 3.8.
	}\label{fig:t:1}
\end{figure}

\begin{figure}
	\centering
	\includegraphics[width=50mm, height=50mm]{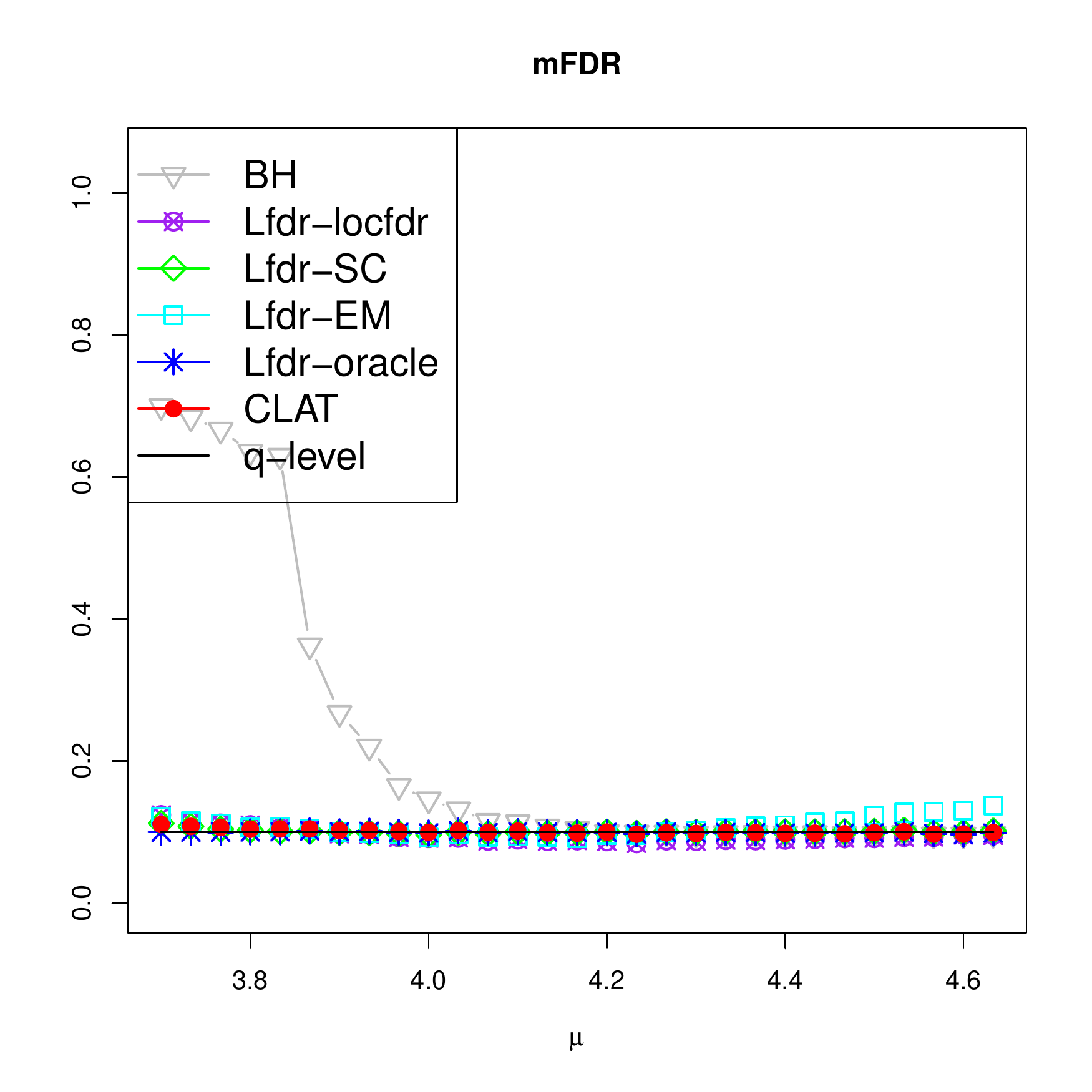}
	\includegraphics[width=50mm, height=50mm]{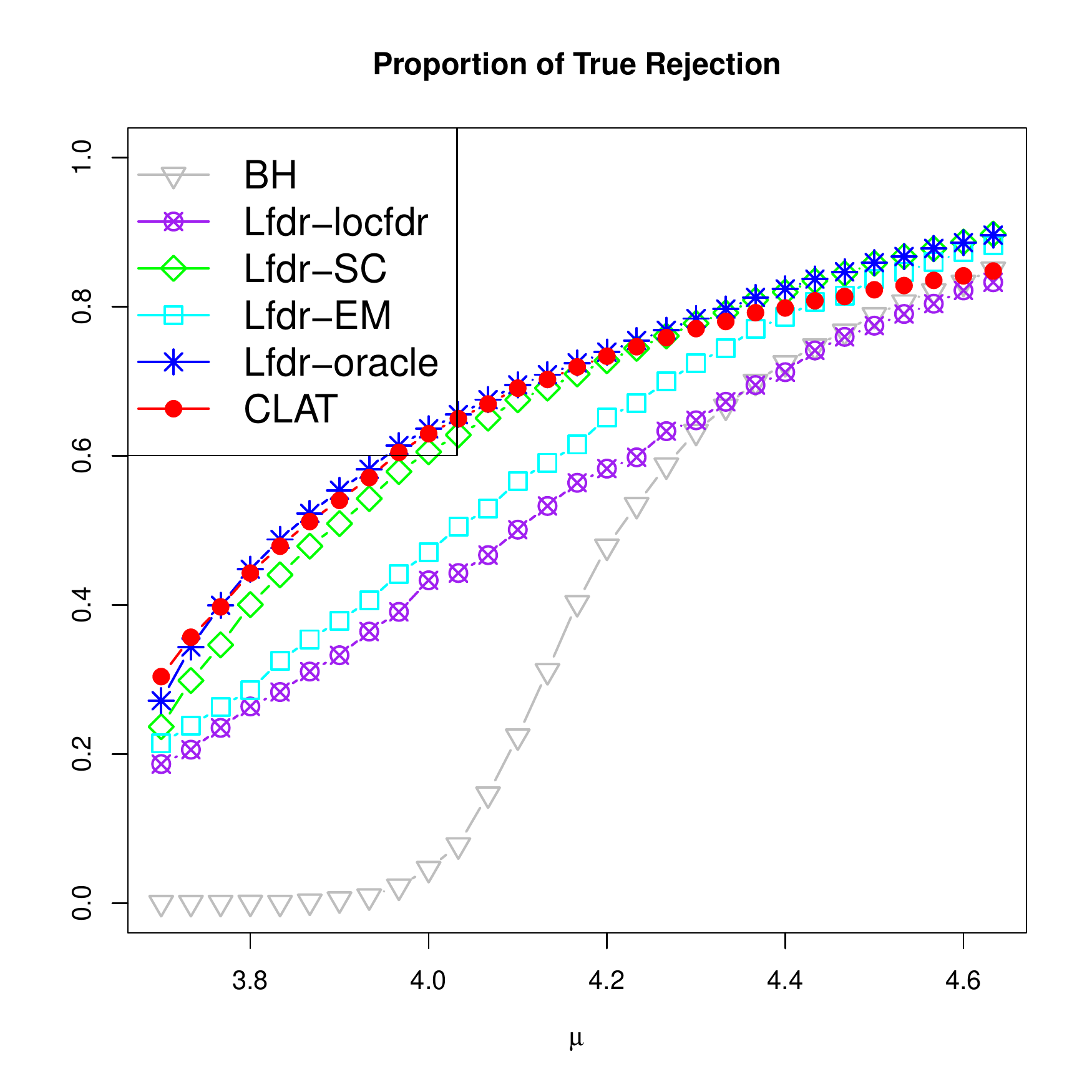}
	\caption{Case II: plot of the \mfdrspace and the average percentages of true rejections. The parameters are chosen as $n=5,000$, $df=10$, $\beta=0.4$, $p_1=0.9$, $\sigma=0.7$ and $\mu$ various from 3.6 to 4.6.
	}\label{fig:t:2}
\end{figure}

\begin{figure}
	\centering
	\includegraphics[width=50mm, height=50mm]{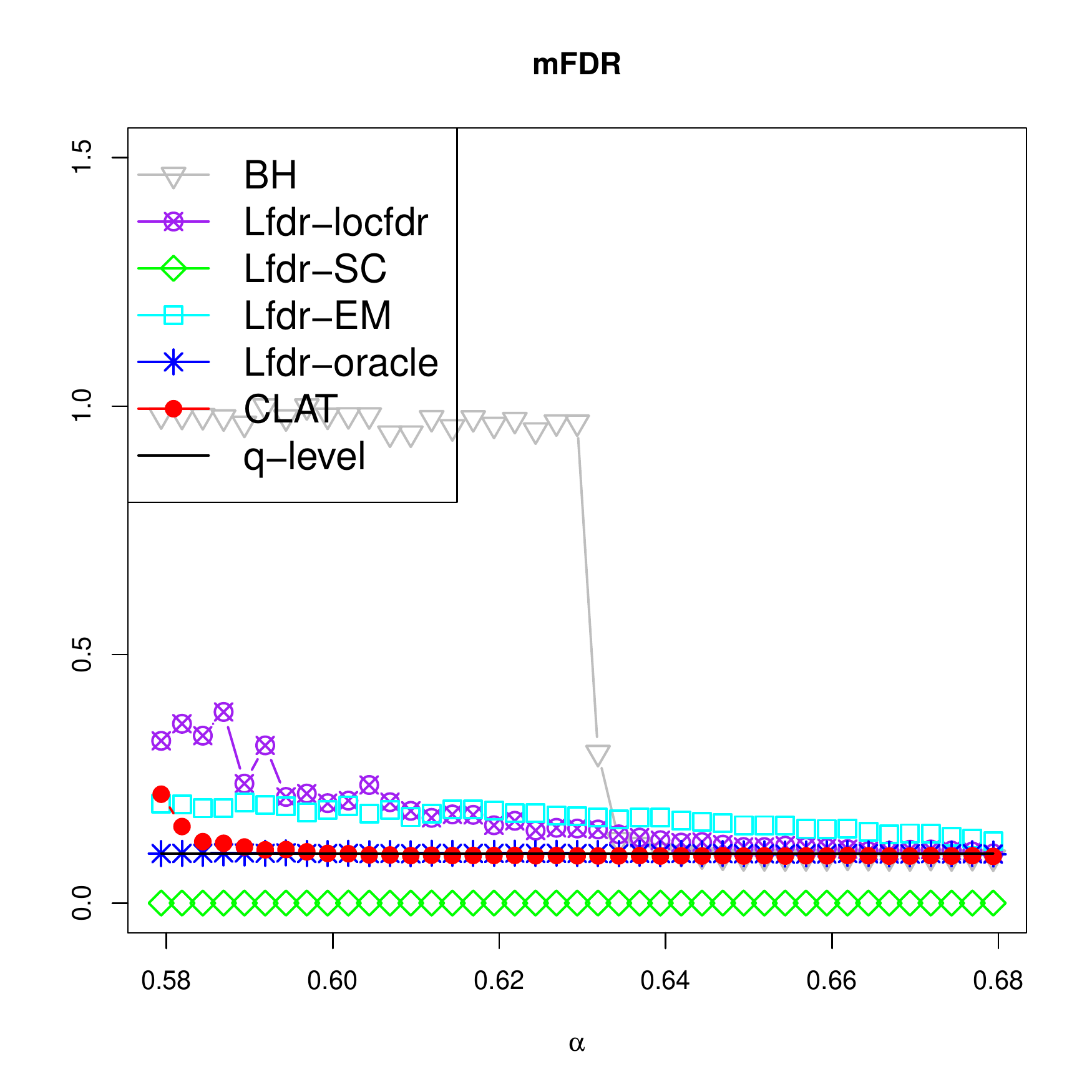}
	\includegraphics[width=50mm, height=50mm]{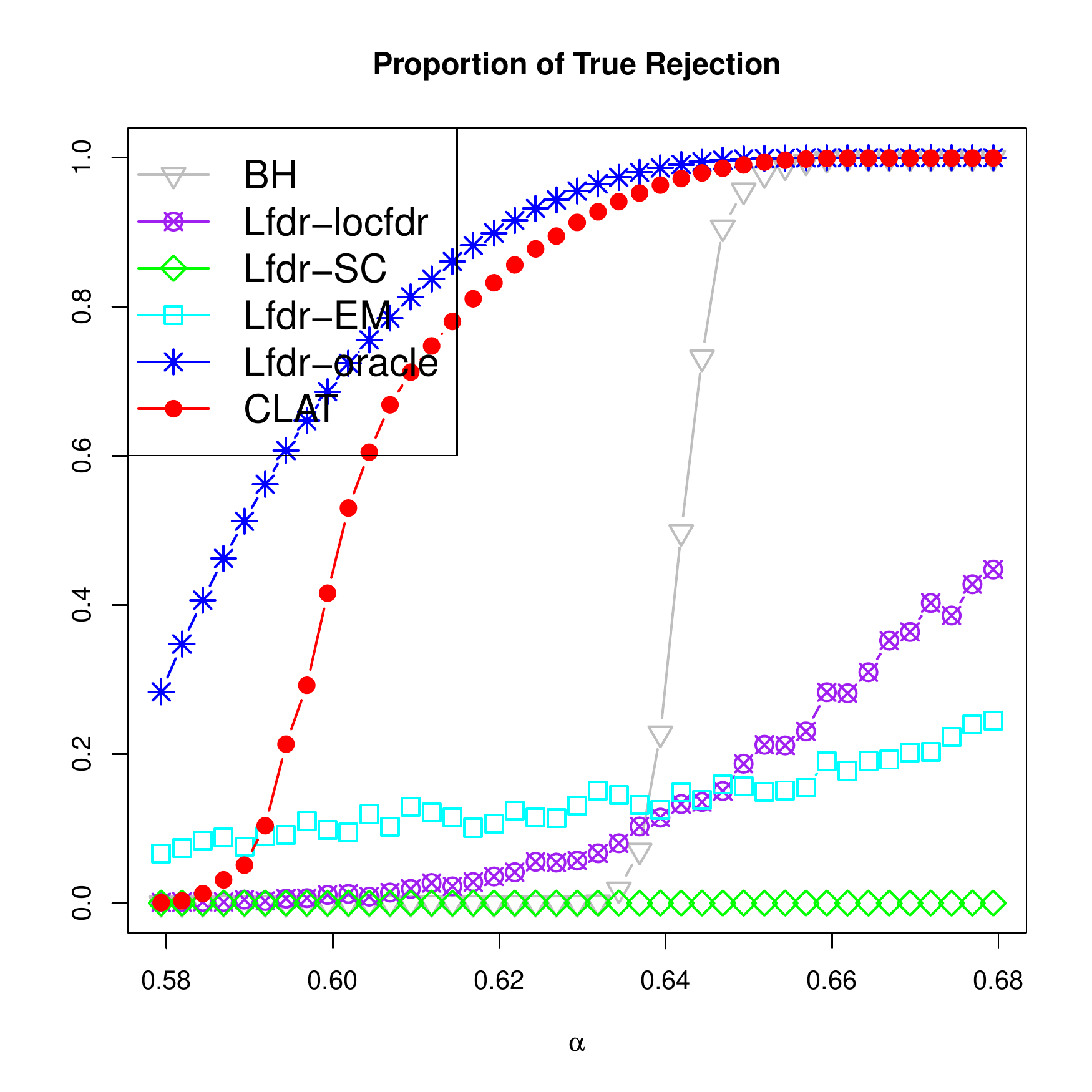}
	\caption{Case III: plot of the \mfdrspace and the average percentages of true rejections. The parameters are chosen as $n=5,000$, $\beta=0.3$, $l=1.2$, and $\alpha$ various from 0.58 to 0.68.
	}\label{fig:unif:1}
\end{figure}

\begin{figure}
	\centering
	\includegraphics[width=50mm, height=50mm]{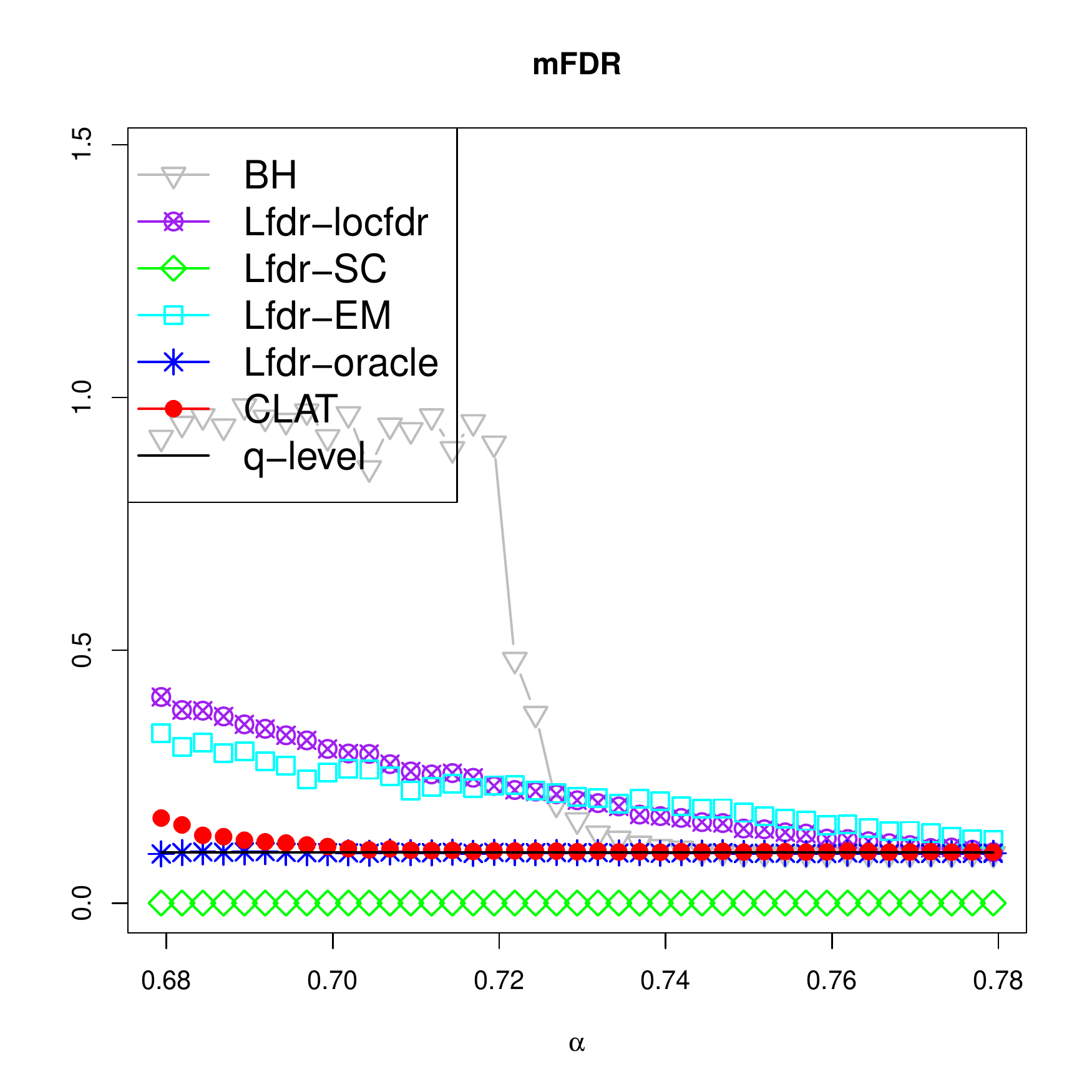}
	\includegraphics[width=50mm, height=50mm]{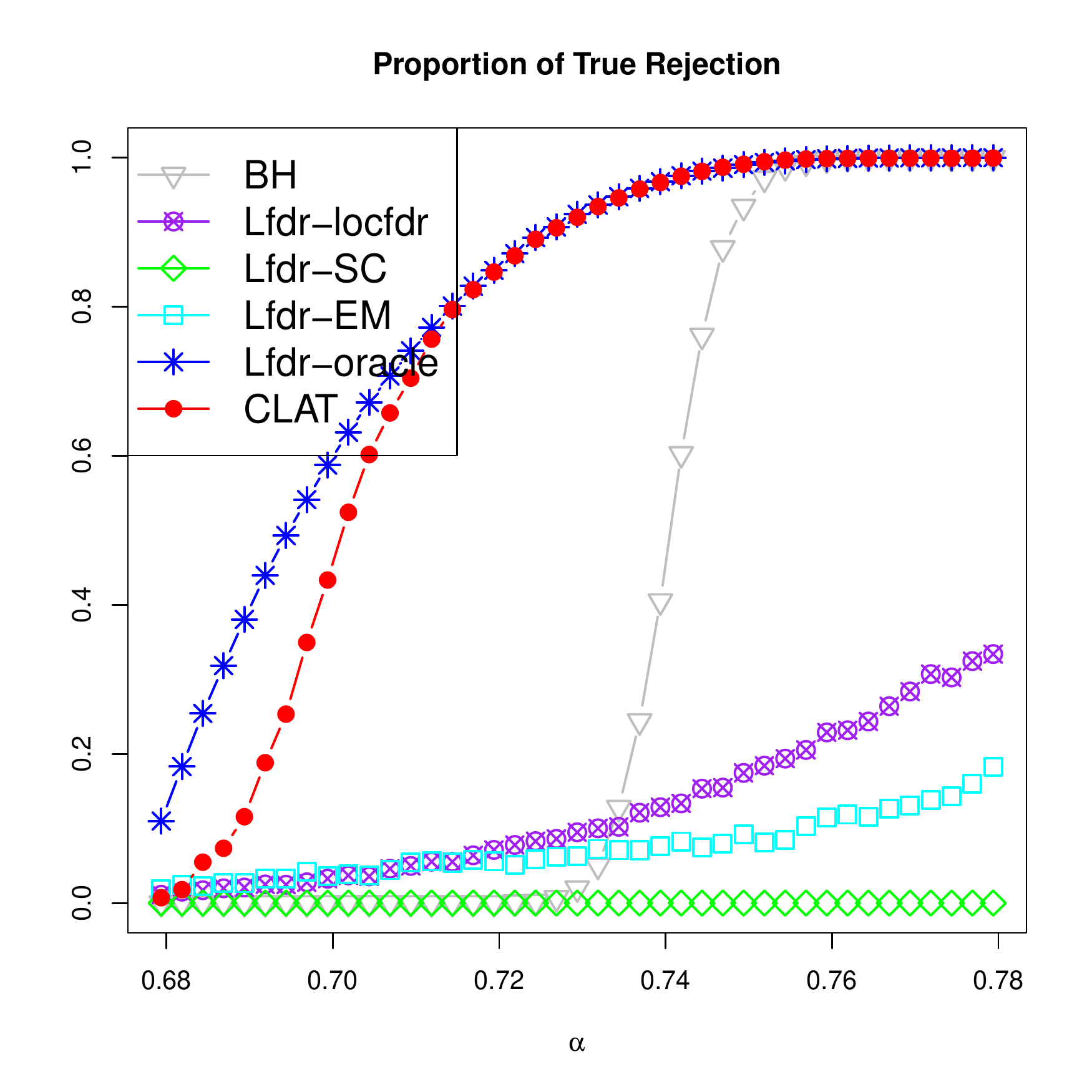}
	\caption{Case III: plot of the \mfdrspace and the average percentages of true rejections. The parameters are chosen as $n=5,000$, $\beta=0.4$, $l=1.2$ and $\alpha$ various from 0.68 to 0.78.
	}\label{fig:unif:2}
\end{figure}

For the local fdr based methods, the data are transformed such that the transformed statistic follows a standard normal distribution under the null hypothesis. For Case II, set $Z_i=\Phi^{-1}(T_d(X_i))$ where $T_d(x)$ and $\Phi(x)$ are the cumulative distribution function of student's t-distribution with degrees of freedom $d$ and the standard normal distribution respectively. For case III, let $Z_i=\Phi^{-1}(1-X_i)$. The number of mixture components $L$ in the EM algorithm are set as 2, 2, 1, and 2 in these four cases. The desired \mfdrspace level $q$ is set as $0.1$.

For a given parameter setting, we replicate the simulations 500 times to calculate \mfdrspace and the average proportion of the number of true rejections over the total number of non-nulls. The results are reported in Figures \ref{fig:normal:1}-\ref{fig:unif:2}. In Case I and Case II, we set $n=5,000, p_1=0.9, \sigma=0.7$, and $\beta$ is chosen as 0.3 and 0.4 respectively. The parameter $\mu$ is chosen such that the maximum likelihood ratio is greater than $q'$, which is the condition specified in Theorem \ref{thm:prop:suff}. In Case III, we set $n=5,000, l=1.2$, and $\beta$ is chosen as 0.3 and 0.4 respectively. The parameter $\alpha$ is chosen such that the maximum likelihood ratio is greater than $q'$.

We call a method valid when the \mfdrspace is less than or equal to the $q$-level for all parameter settings. The Lfdr-oracle is the benchmark. We find that the CLAT, the BH method and Lfdr-SC are valid. For all the cases, the proportion of true rejections for the BH method is substantially smaller than that of CLAT. For Cases I and II, the proportion of true rejections based on Lfdr-SC is similar to that of CLAT. However, the CLAT method rejects a much higher number of hypotheses than the Lfdr-SC for Case III. The Lfdr-EM is not valid and the \mfdrspace could be inflated to a level that is much higher than $q$. One explanation is that when $\pi_1=n^{-\beta}$ decays to zero, it is difficult to obtain a consistent estimator for the parameters. The Lfdr-locfdr also fails to control \mfdrspace under many parameter settings and is not valid.

For Case IV when the test statistics are dependent, the Lfdr-EM is not valid. Both the BH method and Lfdr-locfdr are valid but conservative. The \mfdrspace of the Lfdr-SC method could be slightly higher than the q-level. In contrast, the CLAT is valid and is powerful in rejecting hypotheses. 

\begin{figure}
	\centering
	\includegraphics[width=50mm, height=50mm]{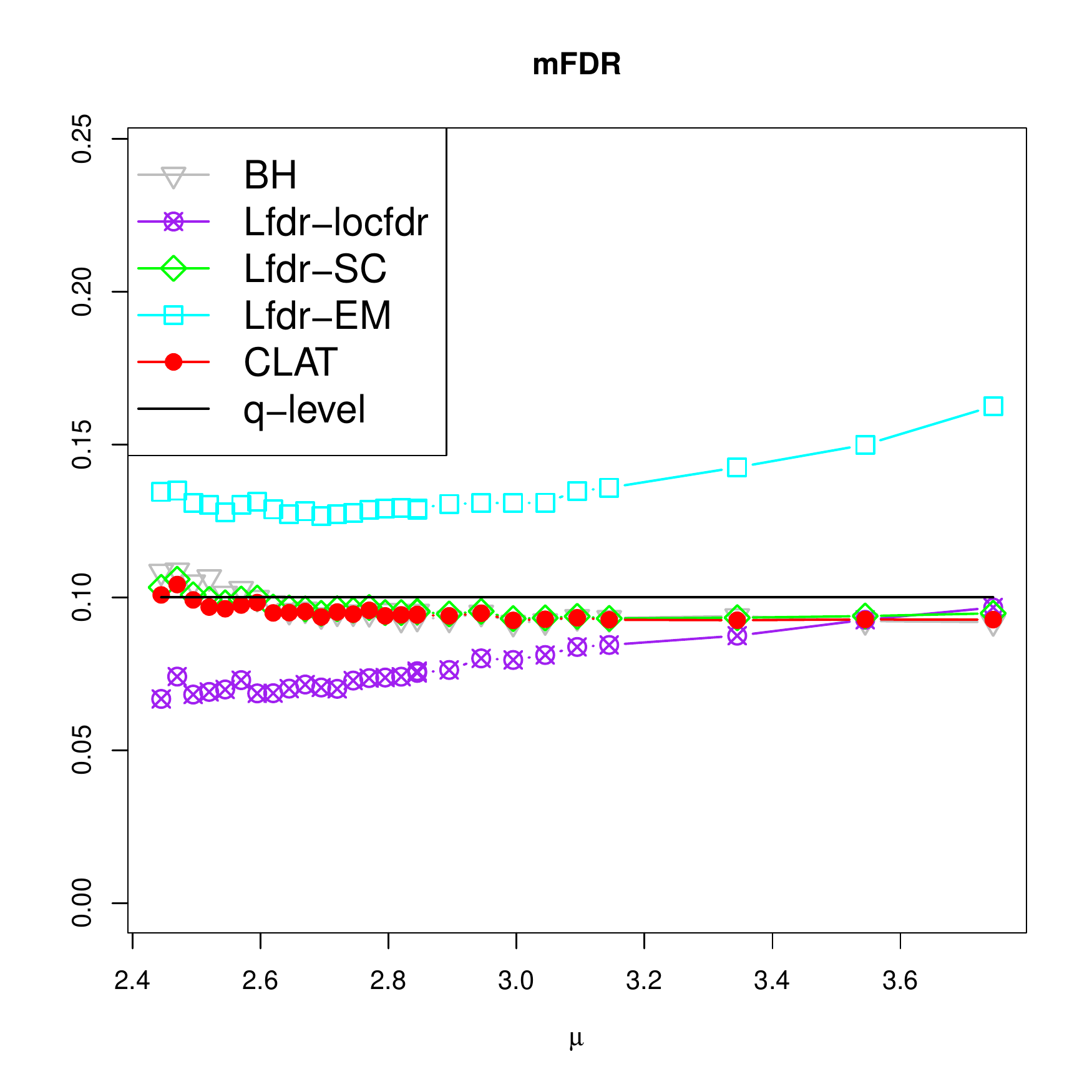}
	\includegraphics[width=50mm, height=50mm]{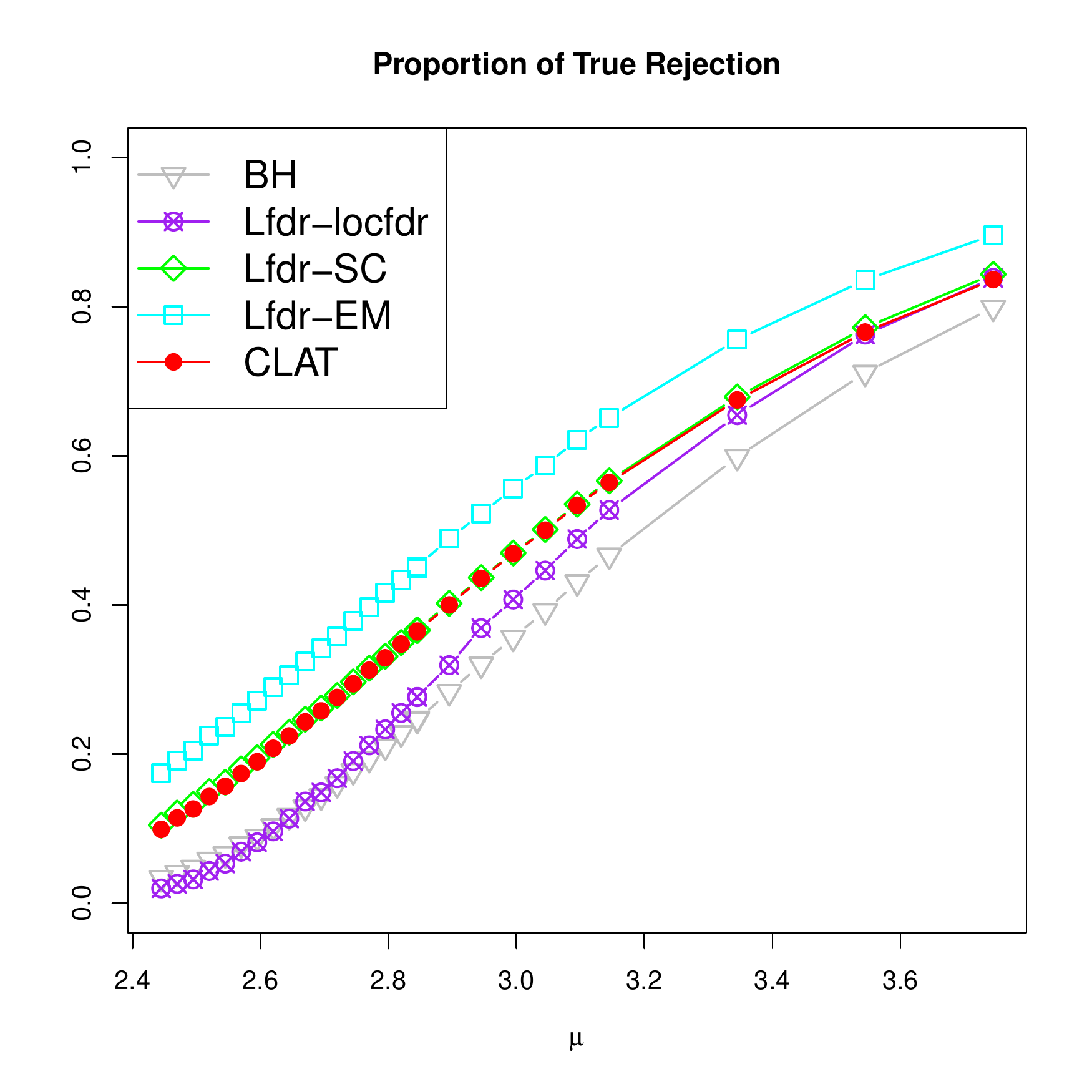}
	\caption{Case IV: plot of the \mfdrspace and the average percentages of true rejections. The parameters are chosen as $n=5,000$, $\beta=0.3$, $p_1=0.9$, $\sigma=0.7$, $\sigma_2=0.5$, and $\mu$ various from 2.4 to 3.8.
	}\label{fig:dep_normal:1}
\end{figure}

\begin{figure}
	\centering
	\includegraphics[width=50mm, height=50mm]{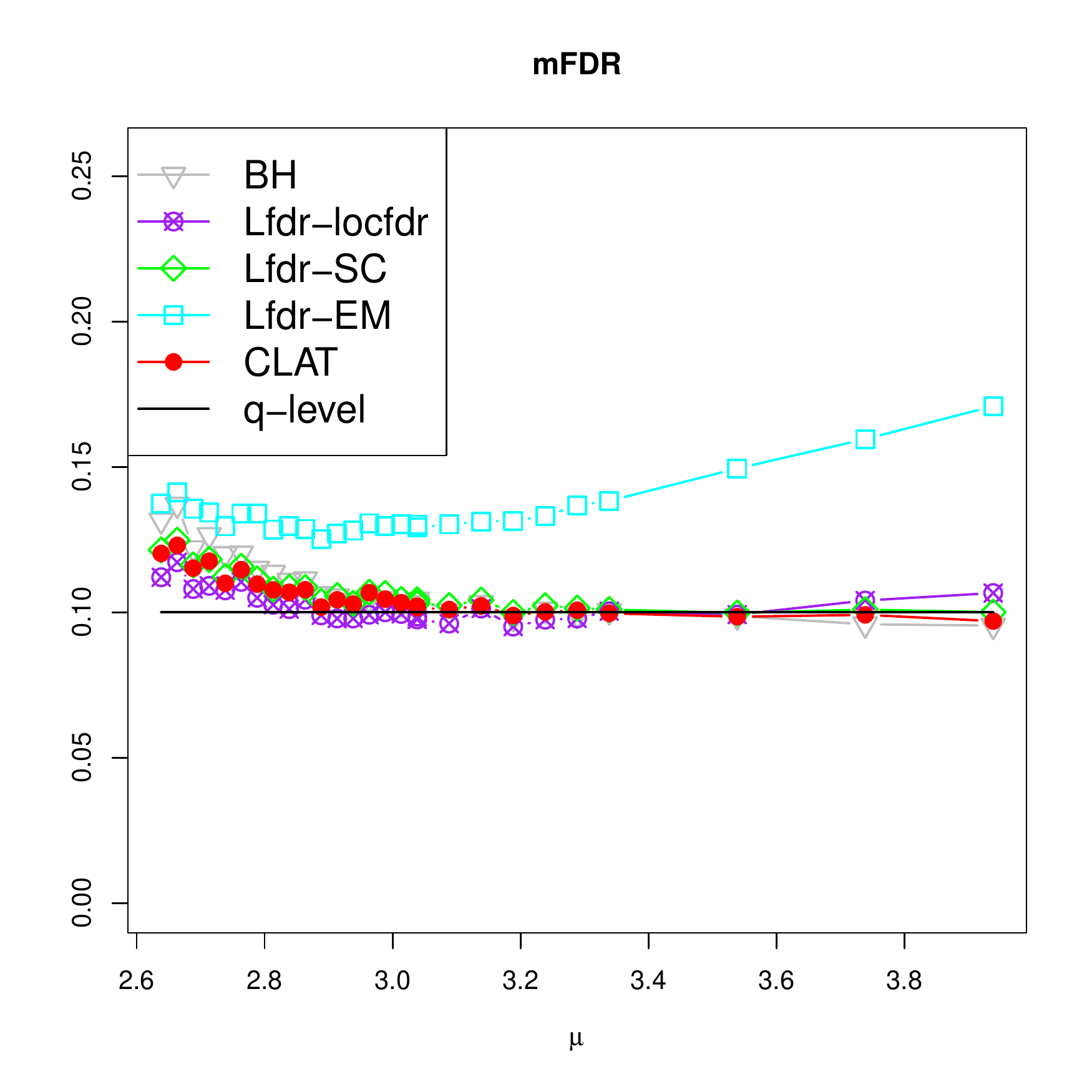}
	\includegraphics[width=50mm,  height=50mm]{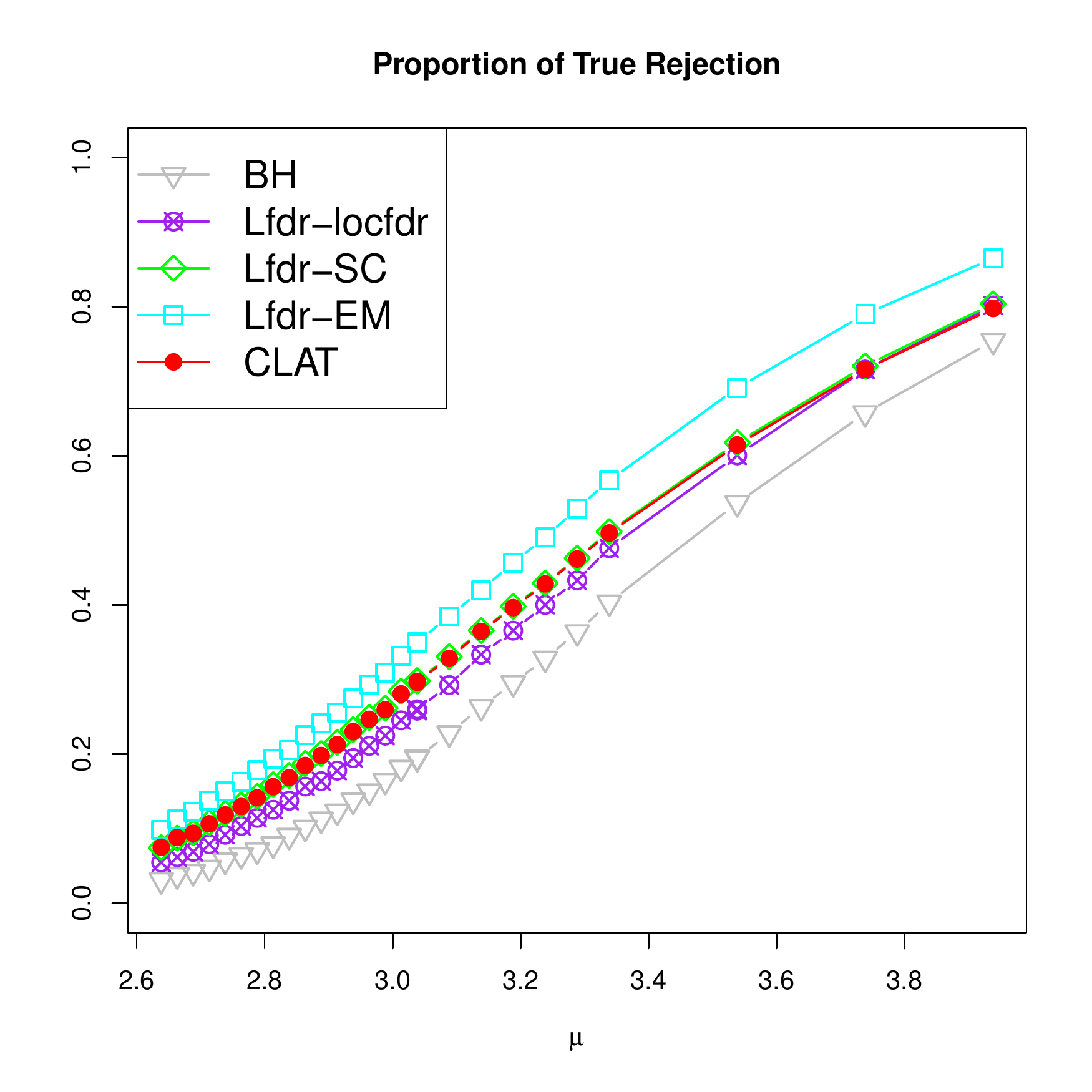}
	\caption{Case IV: plot of the \mfdrspace and the average  percentages of true rejections. The parameters are chosen as $n=5,000$, $\beta=0.4$, $p_1=0.9$, $\sigma=0.7$, $\sigma_2=0.5$, and $\mu$ various from 2.4 to 3.8.
	}\label{fig:dep_normal:1}
\end{figure}

In summary, under all the simulation settings, the CLAT is powerful in rejecting the hypothesis subject to a control of \mfdrspace at the designated level. Competing methods are often either too conservative or liberal depending on the scenario and parameter settings.

\section{Data Analysis}\label{sec:data}
In this section, we apply various procedures to the Golden Spike data set (\cite{Choe:Bouttros:Michelson:Chruch:Halfon:2005}). In this microarray experiment, there are six arrays under two conditions, with three replicates per condition. Among all 14,010 probesets in each array, 1,331 have been spiked-in at higher concentrations in one condition relative to the other. Consequently, this data set has a large number of differentially expressed probesets and a large number of non-differentially expressed probesets. This microarray data set could be used to validate statistical methods (\cite{pearson2008comprehensive}).

We process the data according to \citet{Hwang:Qiu:Zhao:2009}. Let $t_i$ be the $T$ statistic with the degrees of freedom $d_i$ determined by the Satterthwaite approximation. The $Z$-statistic as
$
z_i=\Phi^{-1}(T_{d_i}(t_i)),
$
where $\Phi$ and $T_{d_i}$ are the cumulative distribution functions of the standard normal distribution and the student's t-distribution with $d_i$ degrees of freedom. It is shown in Figure \ref{fig:lr:Golden} that the estimated likelihood ratio $\hat{\Lambda}(x)$ is not monotone.

We then apply different approaches to these $z_i$'s. The \mfdrspace level $q$ we are aiming to control
are set as $0.05$ and $0.10$ respectively. The results are reported in the first two rows of Table \ref{tab:golden:1}. In each cell, we report the number of true rejections and the number of false rejections. The Lfdr-locfdr fails for this data set. There are too many false positives using the Lfdr-EM. The CLAT performs better than the BH method as it yields more true positives and fewer false positives. Lfdr-SC tends to have a larger number of true positives; however, the number of false positive of Lfdr-SC is much greater than that of the CLAT. 

To put them in a fair comparison, we adjust the q-level such that the actual FDPs of various methods are 0.05 and 0.1, and report the average number of true rejections and false rejections in the last two rows of Table \ref{tab:golden:1}. It is shown that the CLAT yields the highest number of true positives than all its competitors. 



\begin{table}
\begin{tabular}{|c|c|c|c|c|c|}
\hline
q&CLAT & BH & Lfdr-SC & Lfdr-locfdr & Lfdr-EM \\
\hline
0.05 & 728/88 & 692/107 & 809/181 & 0/0 & 838/223\\
\hline
0.10 & 859/200 & 760/249 & 914/447 & 0/0 & 973/644\\
\hline
\hline
\multicolumn{6}{|l|}{Set a threshold such that actual FDPs of all methods are 0.05 and 0.1. }\\ 
\hline
0.05 & 543/28 & 168/7 & 521/27 & 0/0 & 515/27 \\
\hline
0.1 & 708/78 & 444/48 & 675/67 & 0/0 & 592/48 \\
\hline
\end{tabular}
\caption{Golden Spike data: this table summaries the data analysis result of five testing procedures when applied to the Gold Spike data set. In each cell, two numbers correspond to the number of
true positives and false positives among all rejections.
}\label{tab:golden:1}
\end{table}

\section{Conclusion}
Testing multiple hypothesis has been an important problem in the last three decades. In this article, we investigate the limitations of some commonly used approaches and propose a new method, the CLAT. We argue that the CLAT has a three-fold advantage over comparable methods: (i) it is optimal for a broader family of distributions; (ii) it is a non-parametric method and relies on the empirical distribution function only; and (iii) it can be computed instantaneously. Both simulations and real data analysis have demonstrated its superiority over other existing methods.

When the MLR holds, the CLAT produces results similar to the BH method. For cases when the MLR does not hold, the CLAT will reject hypotheses with p-values of moderate magnitudes. Namely, the common intuition that we should reject the null when the p-value is smaller than certain threshold is no longer true. The main reason is that the commonly defined p-value relies on the distribution of the test statistic under the null hypothesis only. It fails to use the information of the (unknown) alternative distribution. Under the traditional setting when dealing with a handful of hypotheses, one can not reliably estimate the alternative distribution. However, in the modern applications when often handle thousands or even hundreds of thousands parameters simultaneously, it is possible to obtain a reliable estimator of the alternative distribution which could provide additional insight on choosing a rejection region different from the one based on common intuition. This could lead to better power as demonstrated.

Additionally, when taking another perspective of testing from the Bayesian viewpoint, the decision should depend on the posterior probability that a null hypothesis is true, which is essentially equivalent to the local fdr. Depending on whether the likelihood function is monotonic or not, this posterior probability does not always decrease when the magnitude of the test statistic increases. The CLAT relaxes the requirement of the likelihood function and could be adaptive to the condition of the likelihood ratio.

From the numerical studies, it is shown that the CLAT is valid for dependent data. The argument in the proof of consistency relies on the empirical distribution function. It appears possible to establish theoretical results for the dependence case as long as the empirical distribution function converges to the cumulative distribution function. We will leave this for future research.

The code for CLAT and numerical experiments is available on \url{https://github.com/zhaozhg81/CLAT} and the technical proofs and the EM algorithm are put in the appendix.

\section{Acknowledgement}
This research is supported in part by NSF Grant DMS-1208735 and NSF Grant IIS-1633283. The author is grateful for initial discussions and helpful comments from Dr. Jiashun Jin.

\bibliographystyle{spbasic}      
\bibliography{zhaozhg}

%
%

\newpage
\section{Appendix}\label{sec:app}

\subsection{\bf Proof of Theorem \ref{thm:1}:}
(a) Theorem 2.2 and its proof in \cite{He:Sarkar:Zhao:2015}, the optimal rejection set $\mathbbm{S}_F(q)$ is given as
\[
\mathbbm{S}_F(q) = \{x: \Lambda(x)>c\},
\]
where $c$ is chosen as the minimum value such that \mfdrspace is less than or equal to $q$.

When $\Lambda$ is monotone increasing, then $\mathbbm{S}_F(q) = (c', \infty)$. This agrees with the $\mathbbm{I}_{BH}(q)$ and $\mathbbm{I}_F(q)$ defined in Equation (\ref{eq:gen:bh}).

(b) When $\mathbbm{S}_F(q)$ is a finite interval, by the definition, $\mathbbm{I}_F(q) = \mathbbm{S}_F(q)$. Since the right end point of the interval $\mathbbm{I}_F(q)$ is $\infty$, it is not optimal.
\qed

\subsection{\bf Proof of Theorem \ref{thm:prop:ness}:} 
For any interval $\mathbbm{I}_i=[a,b]$, let
$s(a,b)=(1-\pi_1)\int_a^bdF_0(x)-q\int_a^bdF(x)$. Then
\[
\frac{\partial s}{\partial b}=(1-q)(1-\pi_1)f_0(b)\left(1-\frac{\Lambda(b)}{q'}\right)>0.
\]
Consequently, for any fixed $a$, $s(a, b)$ is increasing with respect to $b$. Since $s(a, a)=0$, therefore, $s(a, b)>0, \forall b>a$. This implies that $(1-\pi_1)\int_{\mathbbm{I}_i}dF_0(x)> q\int_{\mathbbm{I}_i}dF(x)$, for all $i=1,2,\cdots$. As a result,
\[
(1-\pi_1)\int_{\mathbbm{U}}dF_0(x)> q\int_{\mathbbm{U}}dF(x),
\]
 which completes the proof. \qed

\subsection{\bf Proof of Theorem \ref{thm:prop:suff}:} Let $s(a,b)=(1-\pi_1)\int_a^b dF_0(x) - q\int_a^b dF(x)$. Consider $a=c_1$. Then $s(c_1,c_1)=0$. According to the proof of Theorem \ref{thm:prop:ness}, $\frac{\partial s}{\partial b}<0, \forall b\in [c_1,c_2]$. This implies that $s(c_1,c_2)<0$ and consequently $[c_1,c_2]\subset  \mathbbm{S}_F(q)$.

\qed

\subsection{\bf Proof of Theorem \ref{thm:prop:suff2}:}
Define the function $s(a)=(1-\pi_1) \int_a^\infty dF_0(x) - q\int_a^\infty dF(x)$. Then 
\[
s'(a) = -(1-\pi_1)f_0(a) + \pi_1 f(a) = q\pi_1 f_0(a) (\Lambda(a)-q').
\]
Let $c$ be the value such that $\Lambda(c)=q'$. When $a\ge c$, $s'(a)>0$, implying that $s(a)$ is increasing with respect to $a$. Since $s(\infty)=0$, therefore $s(c)<0$. Consequently, $\mathbbm{I}_{F}(q)$ contains $[c,\infty)$.
\qed

\subsection{\bf Proof of Theorem \ref{thm:conRate}:} 
\begin{figure}
  \centering
  \includegraphics[width=50mm, height=50mm]{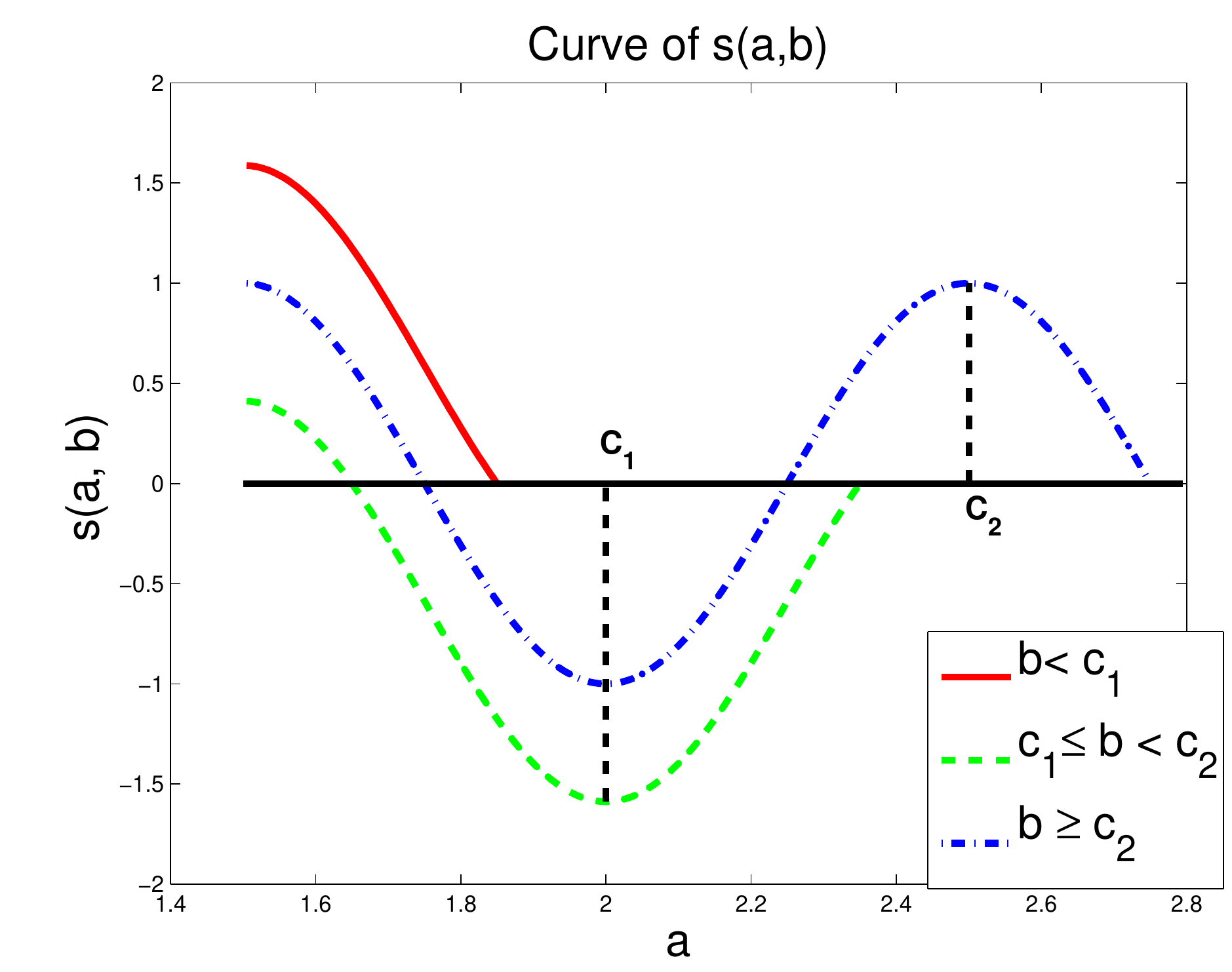}
  \includegraphics[width=50mm, height=50mm]{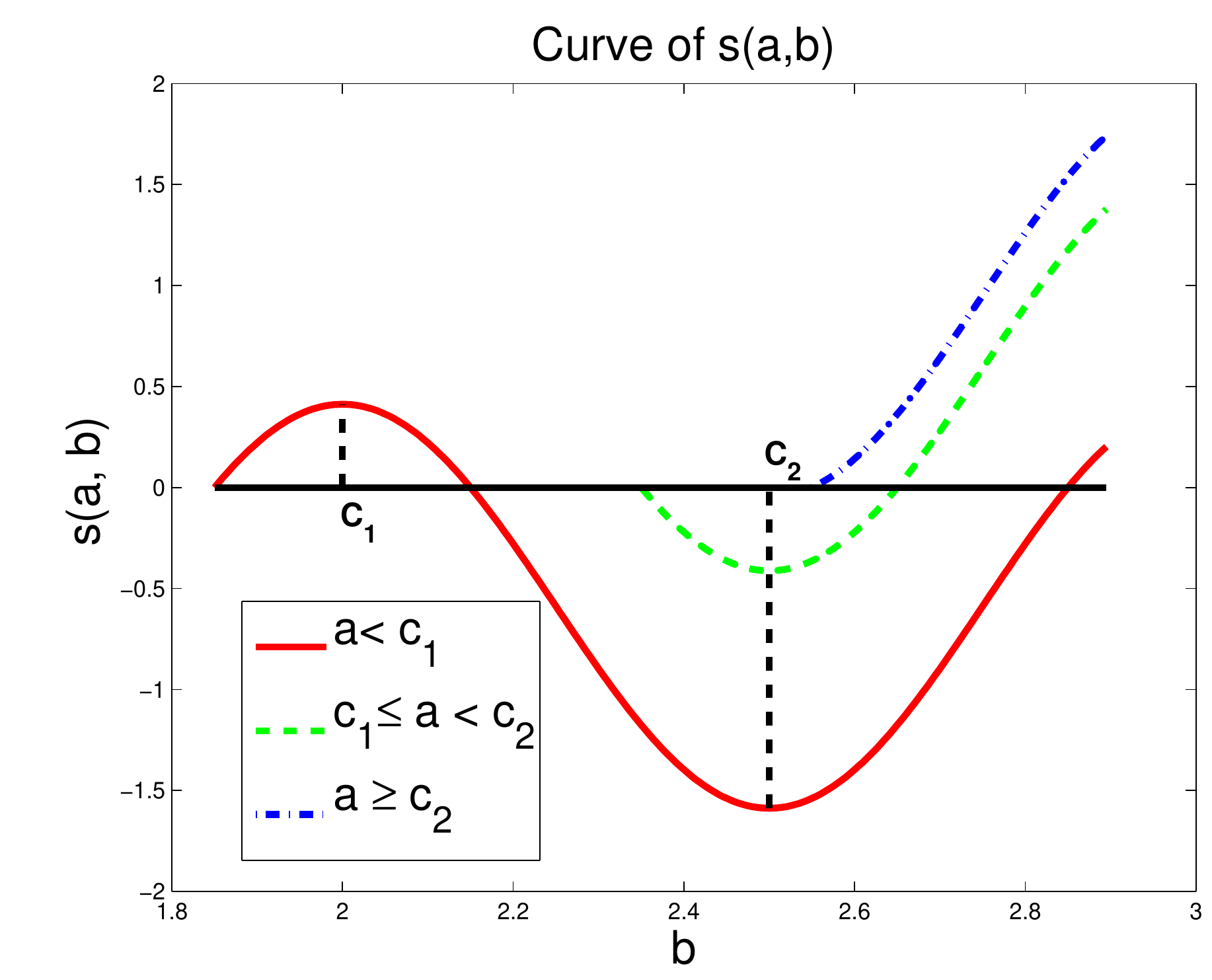}
  \caption{ Curve of the function of $s(a,b)$. In the left panel, $b$ is a fixed constant and we plot $s(a,b)$ as a function of $a$. In the right panel, we plot it as a function of $b$ with $a$ being fixed as a constant.
    }\label{fig:s_a_b}
\end{figure}

According to the definition of $s(a,b)$ and $c_1, c_2$, we know that
\begin{eqnarray*}
\frac{\partial s}{\partial b}=(1-q)(1-\pi_1)f_0(b)\left(1-\frac{1}{q'}\Lambda(b)\right)\left\{\begin{array}{cc}>0, & \textrm{if $b<c_1$},\\ <0, & \textrm{if $c_1<b<c_2$},\\>0, & \textrm{if $b>c_2$}.\end{array}\right.
\end{eqnarray*}
Consequently, for any fixed $a$, $s(a,b)$ increases when $b<c_1$ or $b>c_2$ and decreases when $c_1<b<c_2$.
Similarly,
\begin{eqnarray*}
\frac{\partial s}{\partial a}=(1-q)(1-\pi_1)f_0(a)\left(\frac{1}{q'}\Lambda(a)-1\right)\left\{\begin{array}{cc}<0, & \textrm{if $a<c_1$},\\ >0, & \textrm{if $c_1<a<c_2$},\\<0, & \textrm{if $b>c_2$}.\end{array}\right.
\end{eqnarray*}
For any fixed $b$, $s(a,b)$ decreases when $a<c_1$ or $a>c_2$ and inreases when $c_1<a<c_2$. To demonstrate this pattern, we plot various curves of $s(a,b)$ in Figure \ref{fig:s_a_b}.

Since $g(a)$ attains the maximum at $a_0$, according to Theorem \ref{thm:prop:suff}, $a_0<c_1$ and $b_{a_0}(F)>c_2$. 
Consequently, $(1-\pi_1)f_0(a_0)-qF'(a_0)>0$, and $(1-\pi_1)f_0(b_{a_0}(F))-qF'(b_{a_0}(F))>0$. 
Therefore, the function $b_a(F)$ is a monotone increasing function of $a$ at a small neighborhood of $a_0$.
For a sufficiently small constant $L$ independent of $n$, there exists a neighborhood $A'$ of $b_{a_0}(F)$ such that $f_0(x)-qF'(x)>L$, $\forall x\in A'\cup b^{-1}_{A'}(F)$ where $b^{-1}_{A'}(F)=\{a: b_a(F)\in A'\}$.
Let $A=[a_1,a_2]=b^{-1}_{A'}(F)$ where $a_1<a_0<a_2<c_1$. 
The proof of Theorem \ref{thm:conRate} requires the following lemmas.

\begin{lemma}\label{lem:1}
Let $F_n$ be the empirical distribution function, then $\forall a$, if $b_a(F)=+\infty$ or $b_a(F)<+\infty$ and $F'(b_a(F))-\frac{1}{q}f_0(b_a(F))\neq 0$, then
\begin{equation*}
b_a(F_n)\to b_a(F), \textrm{and}\quad g_n(a)\to g(a).
\end{equation*}
If $F'(b_a(F))-\frac{1}{q}f_0(b_a(F))=0$, then $\limsup g_n(a)\le g(a)$.
\end{lemma}

\begin{lemma}\label{lem:2}
There exists a sub-interval $\mathbbm{B}=[b_1,b_2]$ of $\mathbbm{A}=[a_1,a_2]$, such that for all $a\in \mathbbm{B}$, $|b_a(F_n)-b_a(F)|\le C\epsilon$ provided that $||F_n-F||<\epsilon$.
\end{lemma}

\begin{lemma}\label{lem:3}
The function $g_n(a)$ can not achieve the maximum at $\mathbbm{B}^c$.
\end{lemma}

\begin{lemma}\label{lem:4}
For any $a\in \mathbbm{B}$, $|g_n(a)-g(a)|<C\epsilon$.
\end{lemma}

{\bf Proof of Theorem \ref{thm:conRate}:} Assume that $g_n(a)$ attains the maximum at $a=a_n$, then according to Lemma \ref{lem:3}, $a_n\in \mathbbm{B}$.
According to Lemma \ref{lem:4},
\begin{equation*}
g_n(a_n)-g(a_0)= g_n(a_n)-g_n(a_0)+g_n(a_0)-g(a_0)>-C\epsilon.
\end{equation*}
Since $g(a_n)-g(a_0)<0$, $g_n(a_n)-g(a_0)=g_n(a_n)-g(a_n)+g(a_n)-g(a_0) <C\epsilon$. In other words,
$
|g_n(a_n)-g(a_0)|<C\epsilon.
$
Further, DKW's inequality guarantees that 
$P(\sup_x|F_n(x)-F(x)|>\epsilon)\le 2e^{-2n\epsilon^2}$.
Consequently,
\[
P(|g_n(a_n)-g(a_0)|>C\epsilon)\le 2e^{-2n\epsilon^2}.
\]
Next, we will prove that $\limsup_{n\to\infty} \mfdr\le q$. According to the definition of $a_n$,
\[
\frac{ (1-\pi_1)\int_{a_n}^{b_{a_n}(F_n)} dF_0 }{ g_n(a_n) } = \frac{ (1-\pi_1)\int_{a_n}^{b_{a_n}(F_n)} dF_0 }{  \int_{a_n}^{b_{a_n}(F_n)}dF_n }\le q.
\]
The \mfdrspace can be written as
\[
\mfdr=\frac{ ( 1-\pi_1 )\int_{a_n}^{b_{a_n}(F_n)}dF_0 }{ \int_{a_n}^{b_{a_n}(F_n)}dF}=\frac{ (1-\pi_1)\int_{a_n}^{b_{a_n}(F_n)}dF_0 }{g(a_n)}.
\]
Note that $|g_n(a_n)-g(a_n)|\le |g_n(a_n)-g(a_0)|+|g(a_n)-g(a_0)|\to 0$ and $g(a_n)\to g(a_0)>0$. Consequently,
\[
\limsup_{n\to\infty} \mfdr =\limsup_{n\to\infty} \frac{ (1-\pi_1)\int_{a_n}^{b_{a_n}(F_n)}dF_0 }{g_n(a_n)}\frac{g_n(a_n)}{g(a_n)}\le q.
\]

\noindent {\bf Proof of Lemma \ref{lem:1}:} Since $F_n$ is the empirical cdf, DKW's inequality guarantees that $\forall \epsilon>0$, with high probability
$
F(x)-\epsilon \le F_n \le F(x)+\epsilon ,\forall x.
$
Consider the function
\[
F_U(x)=\left\{\begin{array}{cc} F(x)+\epsilon & \forall x>a \\ F(x)-\epsilon & \forall x\le a\end{array}\right.
\]
Then by the definition of $b_a(F_n)$ and $F_U$,
\begin{equation*}
\frac{1}{q}\le\frac{F_n(b_a(F_n))-F_n(a)}{(1-\pi_1)(F_0(b_a(F_n))-F_0(a))}\le \frac{F_U(b_a(F_n))-F_U(a)}{(1-\pi_1)(F_0(b_a(F_n))-F_0(a))}.
\end{equation*}
Consequently, $b_a(F_n)\le b_a(F_U)$.
Similarly define
\[
F_L(x)=\left\{\begin{array}{cc} F(x)-\epsilon & \forall x>a \\ F(x)+\epsilon & \forall x\le a\end{array}\right.
\]
Then one can similarly show that $b_a(F_L)\le b_a(F_n)$. As a result,
$
b_a(F_L)\le b_a(F_n)\le b_a(F_U).
$
If $(1-\pi_1)f_0(b_a(F))-qF'(b_a(F))\neq 0$ and $b_a(F)<\infty$, then the curve $s(a,b)$
is strictly increasing at a neighbourhood of $b_a(F)$. Consequently, there exists
a neighbourhood $N$ of $b_a(F)$ such that $b_a(F_U)$ and $b_a(F_L)$ fall in
this neighbourhood $N$. Consequently,
$
b_a(F_n)\to b_a(F).
$
If $b_a(F)=+\infty$, then $b_a(F_L)\to \infty$, implying $b_a(F_n)\to b_a(F)$.
Furthermore, 
\begin{eqnarray*}
&&|g_n(a)-g(a)|=|F_n(b_a(F_n))-F_n(a)-F(b_a(F))+F(a)|\\
&\le&|F_n(b_{a}(F_n))-F(b_a(F_n))|+|F(b_a(F_n)-F(b_a(F))|+|F_n(a)-F(a)|\\
&\le&2\epsilon+|F(b_a(F_n)-F(b_a(F))| \to 0.
\end{eqnarray*}

If $(1-\pi_1)f_0(b_a(F))-qF'(b_a(F))=0$, then there exists an neighborhood $C$ of $b_a(F)$ such that $s(a,x)>\delta> 0, \forall x\in C^c\cap [b_a(F), +\infty)$.
Then $b_a(F_n)$ is bounded by $b_a(F_U)$ which converges to $b_a(F)$.
Consequently,
\begin{equation*}
\limsup g_n(a)\le g(a).
\end{equation*}

\noindent {\bf Proof of Lemma \ref{lem:2}:} Let $\mathbbm{B}=[b_1,b_2]$ be a sub-interval of $\mathbbm{A}=[a_1,a_2]$ that contains $a_0$ such that $b_{\mathbbm{B}}(F)\subset b_{\mathbbm{A}}(F)$. For any $a\in \mathbbm{B}$,
let $\Delta= s(a, b_{a_2}(F)) >0$. Since $s(a, b_{a_2}(F))$ is a continuous function of $a$ and $\mathbbm{B}$ is a closed interval,
one can find a common lower bound $\Delta$ such that $s(a, b_{a_2}(F))>\Delta, \forall a\in \mathbbm{B}$.
Since $\frac{\partial s(a, t)}{\partial t}>0$, $\forall t>b_{a_2}(F)$, $s(a, t)>\Delta$ for all $a\in \mathbbm{B}$ and $t>b_{a_2}(F)$. The definition of $b_a(F_n)$ indicates that
\[
(1-\pi_1)( F_0(b_a(F_n))-F_0(a) ) -q ( F_n(b_a(F_n))-F_n(a) )\le 0.
\]
This leads to
\[
( 1-\pi_1) (F_0(b_a(F_n))- F_0(a)) - q (F(b_a(F_n))-F(a)) \le 2q\epsilon <\Delta.
\]
Therefore $b_a(F_n)<b_{a_2}(F)$.

Next, we will show that $b_a(F_n)> b_{a_1}(F)$. According to the definition of $b_a(F)$, $s(a, b_a(F))=0$ and
\begin{equation*}
\frac{\partial s(a,t)}{\partial t}|_{t=b_a(F)}=(1-\pi_1)f_0(b_a(F))-qF'(b_a(F))>0.
\end{equation*}
We can find $t_0<b_a(F), t_0 > b_{a_1}(F)$, such that
\begin{equation*}
(1-\pi_1)(F_0(t_0)-F_0(a))-q(F(t_0)-F(a))=-\Delta<0
\end{equation*}
Therefore for sufficiently small $\epsilon$,
\begin{equation*}
(1-\pi_1)(F_0(t_0)-F_0(a))-q(F_n(t_0)-F_n(a))<-\Delta+2\epsilon<0
\end{equation*}
which implies that $b_a(F_n)>t_0> b_{a_1}(F)$. Consequently, $b_a(F_n)\in b_A(F)$.

Next, we will prove that
$
|b_a(F_n)-b_a(F)|\le L\epsilon.
$
Indeed, since $(1-\pi_1)(F_0(b_a(F_n))-F_0(a))-q(F_n(b_a(F_n))-F_n(a))\le 0$ and 
\begin{equation}\label{eq:2}
(1-\pi_1)(F_0(b_a(F))-F_0(a))-q(F(b_a(F))-F(a))=0,
\end{equation}
then
\begin{equation*}
q(F_n(b_a(F_n))-F(b_a(F)))-(1-\pi_1)(F_0(b_a(F_n))-F_0(b_a(F)))\ge q(F_n(a)-F(a)).
\end{equation*}
As a result,
\begin{eqnarray}\label{eq:3}
&&q(F(b_a(F_n))-F(b_a(F)))-(1-\pi_1)(F_0(b_a(F_n))-F_0(b_a(F)))\nonumber\\
&\ge&q(F_n(a)-F(a))+q(F(b_a(F_n))-F_n(b_a(F_n)))\ge -2q\epsilon.
\end{eqnarray}
By the definition of $b_a(F_n)$, $(1-\pi_1)(F_0(b_a(F_n)^+) - F_0(a))-q(F_n( b_a(F_n)^+) - F_n(a))>0$. With (\ref{eq:2}), we know that
\begin{eqnarray*}
&&q(F( b_a(F_n)^+) - F(b_a(F) ))- (1-\pi_1) ( F_0(b_a(F_n)^+)-F_0(b_a(F))) \\
&<& q( F_n(a)-F(a))+ q (F(b_a(F_n)^+)- F_n(b_a(F_n)^+)) <2q\epsilon.
\end{eqnarray*}
When we take the limit in the previous formula and combine it with (\ref{eq:3}), we see that
\begin{equation*}
| q(F(b_a(F_n))-F(b_a(F)))-(1-\pi_1)(F_0(b_a(F_n))-F_0(b_a(F))) | < 2q\epsilon.
\end{equation*}
Therefore
\begin{equation*}
|(b_a(F_n)-b_a(F))(qF'(\xi)-(1-\pi_1)f_0(\xi))|\le 2q\epsilon.
\end{equation*}
Since $b_a(F), b_a(F_n)\in b_{\mathbbm{A}}(F)$,  $|qF'(\xi)-f_0(\xi)|>L$, we conclude that $|b_a(F_n)-b_a(F)|\le C\epsilon$ for some constant $C$.

\noindent {\bf Proof of Lemma \ref{lem:3}:} Firstly, we will show that there exists a positive constant $\Delta$ such that $g(a_1)-g(a_0)<-\Delta$, $\forall a_1\notin \mathbbm{B}$.

Since
\[
s(-\infty,c_2)=\int_{-\infty}^{c_2}(1-\pi_1)dF_0(x)-q\int_{-\infty}^{c_2}dF(x)>q\pi_1(q'\int_{-\infty}^{c_2}f_0-1)>0,
\]
and $s(a,c_2)$ decreases when $a<c_2$ and increases when $c_1<a<c_2$. Combining this with the fact that $s(c_2,c_2)=0$, one knows that there exists a unique $a^*<c_1$ such that $s(a^*,c_2)=0$.
Let $\mathbbm{I}=\{[a,b]: s(a,b)\le 0\}$ and
\[
\mathbbm{L}=\{a: \textrm{there exists $b>a$ such that $[a,b]\in \mathbbm{I}$}\}.
\]
First, we prove that $\mathbbm{L}=[a^*,c_2)$. Indeed if $a'>c_2$, then for any $b>a'>c_2$,
$
s(a',b)>s(a',a')=0.
$
Iff $a'<a^*<c_1$, then $s(a',b)>s(a^*,b)\ge 0, \forall b>a^*$.
Consequently $\mathbbm{L}\subset [a^*,c_2)$.  On the other hand, for any $a^*\le a \le c_2$, $s(a,c_2)\le s(a^*,c_2)=0$, implying that $[a^*, c_2)\subset \mathbbm{L}$.  Consequently, $\mathbbm{L}=[a^*,c_2)$.
      
      Note that when $c_1<a\le c_2$, $g(a)<g(c_1)$. We thus only need to consider $\mathbbm{L}'=[a^*,c_1]$. The function $g: \mathbbm{L}'\to [0,1]$ is a continuous function and $g(a)$ attains the maximal at a unique point $a=a_0$. Therefore, we can find a positive constant $\Delta$ such that 
\begin{equation*}
g(a_1)-g(a_0)<-\Delta, \forall a_1\in B^c.
\end{equation*}

For any $a_1\in B^c$, 
if $a_1$ satisfies $f_0(b_{a_1}(F))-qF'(b_{a_1}(F))=0$, Lemma \ref{lem:1} implies that
$\limsup_{n\to\infty} g_n(a_1)\le g(a_1)<g(a_0)-\Delta$. The fact that $g_n(a_0)\to g(a_0)$ implies that $g_n(a_1)<g_n(a_0)$ for sufficiently large $n$.

If $(1-\pi_1)f_0(b_{a_1}(F))- qF'(b_{a_1}(F))\neq 0$, then
\begin{eqnarray*}
&&g_n(a_1)-g_n(a_0)=g_n(a_1)-g(a_1)+g(a_1)-g(a_0)+g(a_0)-g_n(a_0) \\
&<& -\Delta+g_n(a_1)-g(a_1)+g(a_0)-g_n(a_0).
\end{eqnarray*}
According to Lemma \ref{lem:1}, $g_n(a_1)\to g(a_1), g_n(a_0)\to g(a)$, then $g_n(a_1)<g_n(a_0)$.
Consequently, $g_n$ attains the maximum in $\mathbbm{B}$.

\noindent {\bf Proof of Lemma \ref{lem:4}:} \begin{eqnarray*}
&&|g_n(a)-g(a)| =|F_n(b_a(F_n))-F_n(a)-F(b_a(F))+F(a)|\\
&=&|F_n(b_a(F_n))-F(b_a(F_n))+ F(b_a(F_n))-F(b_a(F))-(F_n(a)-F(a))|\\
&\le&2\epsilon +|F(b_a(F_n))-F(b_a(F))|\le 2\epsilon+|b_a(F_n)-b_a(F)||F'(\xi)|.
\end{eqnarray*}
According to Lemma \ref{lem:2}, $b_a(F_n)-b_a(F)=O(\epsilon)$, consequently,
$
|g_n(a)-g(a)|\le C\epsilon.
$

\subsection{\bf EM Algorithm.} In this section, we outline the steps of EM algorithm. Let $X_1,X_2,\cdots,X_n$ be the test statistic. We fit the following model
\[
X_i\iid (1-\pi_1)\phi(x) + \pi_1\sum_{l=1}^L p_l\frac{1}{\sigma_l}\phi(\frac{x-\mu_l}{\sigma_l}).
\]
The parameters to be estimated are $\pi_1$, $p_l$, $\mu_l$, and $\sigma_l^2$, for $l=1,2,\cdots, L$.

\begin{algorithm}[H]
	\caption{EM algorithm.  \label{alg:em}}
	\begin{algorithmic}[1]
		\vspace*{1mm}
		\item Set the initial value, $\pi_1^0=0.5$, $p_l^0=\frac{1}{L}$, $\mu_l^0=(-1)^l$, $(\sigma_l^0)^2=1$, $Diff=1$; 
		\item while $Diff > \delta (\textrm{=0.001 by default})$:
		\begin{itemize}
			\item[(a)] Calculate 
			\[
			fdr_i^t(\vx) = \frac{ (1-\pi_1^t)\phi(x_i)}{ (1-\pi_1^t)\phi(x_i) + \pi_1^t\sum_{l=1}^L p_l^t\frac{1}{\sigma_l^t}\phi(\frac{x_i-\mu_l^t}{\sigma_l^t}) };
			\]
			\item[(b)] Calculate 
			\[
			\omega_i^l = \frac{p_l^t\frac{1}{\sigma_l^t}\phi(\frac{x_i-\mu_l^t}{\sigma_l^t})}{\sum_{l=1}^L p_l^t\frac{1}{\sigma_l^t}\phi(\frac{x_i-\mu_l^t}{\sigma_l^t}) }
			\]
			\item[(c)] Update the parameters:
			\[
			\pi_1^{t+1}=\frac{1}{n}\sum_i (1-fdr_i^t(\vx)), p_l^{t+1}=\frac{\sum_i\omega_i^l(1-fdr_i^t(\vx))}{\sum_i(1-fdr_i^t(\vx))},
			\]
			and
			\[
			\mu_l^{t+1} =\frac{\sum_i (1-fdr_i^t(\vx))\omega_i^lx_i }{\sum_i(1-fdr_i^t(\vx))},
						(\sigma_l^2)^{t+1} =\frac{\sum_i (1-fdr_i^t(\vx))\omega_i^l(x_i-\mu_l^{t+1})^2} {\sum_i(1-fdr_i^t(\vx))},
			\]
			\item[(d)] Calculate 
			\[
			Diff = (\pi_1^{t+1}-\pi_1^t)^2+\sum_{l=1}^L \left[ (p_l^{t+1}-p_l^t)^2 + (\mu_l^{t+1}-\mu_l^t)^2 + ((\sigma_l^2)^{t+1}-(\sigma_l^2)^t)^2
			\right];
			\]
		\end{itemize}
		\item Upon convergence, return the estimated parameters.
	\end{algorithmic}
\end{algorithm}

\end{document}